\documentclass[9pt]{IEEEtran}

\usepackage{color}
\usepackage{docmute}
\usepackage{graphicx}
\usepackage{subfig}
\usepackage{MathDefs}
\usepackage{csvsimple}
\usepackage{epstopdf}
\usepackage{amsmath,amsfonts,amssymb,amsthm,pifont}
\usepackage{cite}
\usepackage{hyperref}
\hypersetup{
  colorlinks   = true, 
  urlcolor     = blue, 
  linkcolor    = blue, 
  citecolor = blue,
}

\newtheorem{remark}{Remark}

\usepackage{tabularx}
\newcolumntype{L}[1]{>{\raggedright\let\newline\\\arraybackslash\hspace{0pt}}m{#1}}
\newcolumntype{C}[1]{>{\centering\let\newline\\\arraybackslash\hspace{0pt}}m{#1}}
\newcolumntype{R}[1]{>{\raggedleft\let\newline\\\arraybackslash\hspace{0pt}}m{#1}}

\usepackage{algorithm}
\usepackage{algpseudocode} 

\newenvironment{abbreviations}{\begin{list}{}{}}{\end{list}}

\newcommand{\blue}{ \textcolor{blue} }

\begin{document}
\title{On the Identification of Electrical Equivalent Circuit Models Based on Noisy Measurements}

\author{
Balakumar Balasingam${^\star}{^\dagger}$, {\em Senior Member, IEEE},
and Krishna Pattipati$^\ddagger$, {\em Fellow, IEEE}
\thanks{Submitted to IEEE Transactions on Instrumentation and Measurement, Dec. 2020.}
\thanks{$^\star$ Balakumar Balasingam is the corresponding author at singam@uwindsor.ca.}
\thanks{$^\dagger$Department of Electrical and Computer Engineering, University of Windsor, 401 Sunset Ave., Office\#3051, Windsor, ON N9B3P4, Canada, TP: +1(519) 253-3000 ext. 5431, E-mail: singam@uwindsor.ca.}
\thanks{$^\ddagger$Department of Electrical and Computer Engineering, University of Connecticut, 371 Fairfield Rd, Office\#350, Storrs, CT 06269, USA, TP: +1(860) 486-2890, E-mail: krishna.pattipati@uconn.edu.}
}

\maketitle

\begin{abstract}
Real-time identification of electrical equivalent circuit models is a critical requirement in many practical systems, 
such as batteries and electric motors. 
Significant work has been done in the past developing different types of algorithms for system identification using reduced equivalent circuit models. 
However, little work was done in analyzing the theoretical performance bounds of these approaches. 
Proper understanding of theoretical bounds will help in designing a system that is economical in cost and robust in performance.
In this paper, we analyze the performance of a linear recursive least squares approach to equivalent circuit model identification and show that the least squares approach is both unbiased and efficient when the signal-to-noise ratio is high enough. 
However, we show that, when the signal-to-noise ratio is low -- resembling the case in many practical applications -- the least squares estimator becomes significantly biased. 
Consequently, we develop a parameter estimation approach based on total least squares method and show it to be asymptotically unbiased and efficient at practically low signal-to-noise ratio regions.
Further, we develop a recursive implementation of the total least square algorithm and find it to be slow to converge; for this, we employ a Kalman filter to improve the convergence speed of the total least squares method.
The resulting {\em total Kalman filter} is shown to be both unbiased and efficient in equivalent circuit model parameter identification.
The performance of this filter is analyzed using real-world current profile under fluctuating signal-to-noise ratios. 
Finally, the applicability of the algorithms and analysis in this paper in identifying higher order electrical equivalent circuit models is explained. 
\end{abstract}


\section{Introduction}
Electrical equivalent circuit model (ECM) parameter estimation is a crucial element in battery management systems that help power 
most of today's consumer electronic devices,
aerospace equipment, 
and a fast-growing number of electric vehicles. 
Accurate estimation of ECM parameters helps to execute important battery management functionalities, such as, 
state of charge estimation, 
remaining power estimation and the time to end-of-charge prediction. 
Research in to ECM identification is very active in the recent past; however, existing works significantly lack theoretical insights supported by the Estimation Theory.  
The present paper seeks to contribute to filling this gap. 

\subsection{Literature Review}

Based on what measurements are taken, the ECM identification methods can be categorized into {\em active} and {\em passive} as follows:
\begin{itemize}
\item {\em Active estimation.}
In active estimation, an {\em excitation current} is applied to the battery and the resulting voltage is measured for parameter estimation. 
Active estimation allows one to excite the battery using specially designed waveforms for accurate system identification.
In batteries, active ECM estimation can be done during charging, where the chargers have the opportunity to apply custom excitation currents that are superimposed to the charging current. 
\item {\em Passive estimation.}
The absence of a special excitation current defines passive ECM parameter estimation.
During passive estimation, both the current through the battery and the voltage across it are measured. 
An example of passive estimation is the ECM parameter identification of an electric vehicle battery while the vehicle is on the move; here, both the (regenerative) charging current and the discharging current cannot be predetermined --- these need to be measured. 
Passive ECM estimation is crucial in an electric vehicle in order to accurately estimate important states of the battery pack, such as, state of charge and remaining time until shut down. 
Since passive approaches do not require any special excitation signals or conditions, they fall under the category of {\em real-time estimation of ECM parameters}.  
\end{itemize}

Another way to categorize ECM identification is based on the domain in which the estimation is performed: 
\begin{itemize}
\item {\em Frequency-domain approaches.}
Frequency domain approach is well suited for active estimation where, usually, a sinusoidal signal is applied and the frequency response (impedance) is measured; this is repeated at different frequencies in order to compute a Nyquist plot \cite{waag2013experimental} of the battery impedance; the parameters of the ECM can be inferred based on this plot.   
The frequency domain approach is widely known as electrochemical impedance spectroscopy (EIS). 
Frequency domain approaches are suited only for active estimation, where a predetermined current applied as system excitation signal. 
{In the} literature, EIS based approaches have been widely used in laboratory setting for ECM model identification \cite{marongiu2015influence}
and for state of health modeling \cite{mingant2016novel}. 

\item {\em Time-domain approaches.}
The advantage of the time-domain approach over its frequency domain counterpart is that the time-domain methods are suitable for real-time parameter estimation due to the fact that they do not need an excitation signal in a specific form. 
Time domain approaches use opportunistic voltage and current measurements in order to estimate the ECM parameters \cite{BFGpart12014,he2012online}. 

\end{itemize}

Existing approaches used several reduced order models, such as,
R-int model, 
RC model, 
Thevenin model, 
Enhanced Thevenin model, 
PNGV model and 
various combined models based on the above circuits \cite{nikdel2014various}. 
Several estimation approaches were employed in the past, 
from linear estimators such as the least square estimator \cite{BFGpart12014} and Kalman filter to non-linear estimators, such as the extended Kalman filter, unscented Kalman filter and particle filter. 
Most of the existing approaches also considered ECM parameter identification as a joint estimation problem with the state of charge estimation. 
Regardless of the numerous approaches presented in the literature for ECM parameter estimation, little was reported about the theoretical performance analyses of these approaches.

The focus of this paper is on time-domain approaches to ECM parameter identification; particularly, we focus on theoretical analyses that inform the performance bounds of ECM identification techniques. 
When it comes to parameter estimation from linear models, such as the R-int model, theoretical performance bounds convey crucial information about the performance of the estimation method;
a model that is proved to be {\em unbiased} will yield zero estimation error in an average sense; and 
a model that is proved to be {\em efficient} will have variance of the estimation error equal to the theoretically derived Cramer-Rao lower bound (CRLB) \cite{bar2004estimation}. 
Theoretical analysis of battery state identification techniques is very sparse in the literature. 
Table \ref{table:performAnalysis} summarizes the existing work on time-domain theoretical performance analysis. 
\begin{table}[h]
\begin{center}
\caption{Time-domain approaches to ECM identification}
\label{table:performAnalysis}
\begin{tabular}{|c|c|c|c|}
\hline
Paper & Active/Passive?   & Current excitation   \\
\hline
\cite{lin2015analytic} & Passive    & square wave \\
\hline
\cite{song2018parameter} & Passive    & sinusoidal \\
\hline
\cite{song2019combined} & Passive    & sinusoidal \\
\hline
\cite{song2018parameterCRLB} & Passive    & sinusoidal \\
\hline
\cite{klintberg2017theoretical} & Passive   & sinusoidal \\
\hline
\cite{lin2017analytic} & Passive    & sinusoidal\\
\hline
\cite{lin2018theoretical} & Active    & generic \\
\hline
\cite{bizeray2018identifiability} & Passive    & sinusoidal \\ \hline
\cite{sharma2014fisher} &  Passive  & generic    \\ \hline
\cite{alavi2016identifiability} &   Passive    & sinusoidal  \\ \hline
\end{tabular}
\end{center}
\end{table}

In addition to the works on theoretical performance analysis summarized in Table \ref{table:performAnalysis}, some other works in the literature reported the importance of systematic performance analysis. 
In \cite{lee2019estimation}, the CRLB was derived for the estimation of relevant parameters of the negative and positive electrode. 
In \cite{ramadesigan2011parameter}, a Gaussian modeling of several parameters and an approach to capacity fade analysis is presented. 
Based on single-particle approaches, a lumped-parameter nonlinear model is developed in \cite{schmidt2010experiment}. 
Using Genetic algorithms, data oriented models and approaches to parameter estimation were presented in \cite{forman2012genetic}. 
A model fitting based and parameter sensitivity analysis were presented in \cite{vazquez2014rapid}. 
In \cite{rothenberger2014maximizing}, an optimization based approach to parameter estimation was presented and the performance bound was derived. 

Compared to the amount of work reporting approaches to battery parameter estimation for use in battery management systems, the reported work on theoretical performance analysis of the estimators is negligible. 
This paper seeks to fill this gap by introducing a number of contributions to the literature as detailed in the next subsection.

\subsection{Summary of Contributions}
The contributions of this paper are summarized below:

\begin{itemize}

 \item {\em Theoretical performance bound of equivalent circuit model identification.}
Theoretical bound on the estimation error variance, the Cramer-Rao lower bound (CRLB), is derived for equivalent circuit model (ECM) parameter identification of a simple R-int model. 
The CRLB can {be} used to {\em stress test} the existing ECM identification methods at various operating conditions. 
Such an analysis will help in designing efficient and robust systems. 

 \item {\em Rigorous performance analysis approach to existing ECM identification methods.}
Performance of traditional ECM parameter estimators were analyzed in terms of {\em bias} and {\em efficiency}. 
Good estimator will have zero bias and it is considered efficient when its estimation error (co)variance is equal to the CRLB. 
Using this approach, the deficiencies in the classical least square based ECM parameter estimator is quantified.
Better estimators are proposed to fill the gaps found using the proposed performance analysis method.

\item {\em Introduction to the notion of signal-to-noise ratio.}
An ECM parameter estimation method, working perfectly at a specific operating condition, might completely collapse in a slightly different operating scenario. 
In this paper, we demonstrate the importance of algorithm evaluation at different signal-to-noise ratios (SNRs).
It is shown that the traditional least square estimator for ECM identification collapses at low SNR region. 
With such an understanding, it is shown that alternative approaches can be developed to remedy this problem.

\item {\em Performance analysis using normalized models.}
The problem of ECM identification arises in wide range of applications. 
For example, in the case rechargeable batteries, a battery management system is a crucial component in small scale applications, such as consumer electronics, as well as in large scale applications, such as electric vehicles. 
The performance analysis presented in this paper is designed in a way that the results are meaningful to devices of any size regardless of their power rating. 
In order to achieve this, the performance measures are normalized to their true (or nominal) values and the analysis is carried out for various SNR regions. 
The reported finings are meaningful regardless of the power rating and size of the application.

\item {\em Time-varying SNR and convergence analysis.}
We showed that the performance of a parameter estimator (defined in terms of the bias and variance of the estimates) depends on the SNR.
This allows one to decide the power of the excitation signal such that a desired performance goal will be satisfied during ECM parameter estimation. 
However, when the ECM parameter needs to be estimated in a passive mode, i.e., in the absence of the luxury of sending an excitation signal, the ECM parameter estimation algorithm is faced with a fluctuating SNR. 
It is demonstrated that the SNR can easily fluctuate from a very favorable region to an unfavorable region. 
This paper provides insights as to what to expect in these situations and how the algorithm should be tuned.

\item {\em Improved parameter estimator using the total Kalman filter.}
Finally, a total Kalman filter (TKF) is introduced as a better approach to ECM parameter estimation in challenging scenarios.

\end{itemize}

\subsection{Paper Organization}

The remainder of this paper is organized as follows:
Section \ref{sec:LS_ideal} gives a preview of parameter estimation in electrical equivalent circuit models by exciting them with current signals. 
This section also summarizes the least squares approach for parameter estimation and demonstrates the performance of the LS estimator in terms of bias and efficiency under different signal-to-noise ratio (SNR) scenarios. 
Section \ref{sec:NoisyModel} illustrates the performance setbacks of the LS estimator when the system model is noisy. 
It further establishes the ECM parameter estimation problem as belonging to the category of systems with noisy observation models. 
Finally, this section introduces the total least squares estimators (TLS) as the most suitable estimator for practical ECM identification. 
Section \ref{sec:RecusiveEST} introduces recursive parameter estimation and shows the trade-off of the TLS estimator in terms of estimator variance and convergence time.
In order to remedy the slow convergence problem of TLS, a total Kalman filter (TKF) is proposed and demonstrated in this section. 
Section \ref{sec:time-var-snr} analyzes the performance of the recursive filters in a realistic situation where the SNR of the system rapidly fluctuates between favourable and unfavourable conditions. 
Section \ref{sec:applications} presents a discussion of the applications of the approaches presented in this paper in practical systems, particularly in rechargeable batteries. 
Finally, the paper is concluded in Section \ref{sec:conclusions}. 

\section*{List of Acronyms}
\begin{abbreviations}
\item[ECM] Equivalent circuit model
\item[EIS] Electrochemical impedance spectroscopy
\item[LS] Least squares
\item[RLS] Recursive least squares
\item[TLS] Total least squares
\item[TKF] Total Kalman filter
\item[SDE] Standard deviation of error
\item[SNR] Signal to noise ratio
\item[CRLB] Cramer-Rao lower bound
\end{abbreviations}

\section*{List of Notations} 
\begin{abbreviations}
\item[$\bA_\kappa$] Measurement model \eqref{eq:vk}
\item[$b_{\rm LS} $] Bias of the LS estimator of $R$ \eqref{eq:muLS}
\item[$\bar b_{\rm LS}$] Normalized bias of the LS estimator of $R$ \eqref{eq:NORMbias}
\item[$\bb$] ECM parameter vector \eqref{eq:vk}
\item[$\hat \bb_{\rm LS} $] LS estimate of $\bb$ \eqref{eq:bkhat}
\item[$\hat \bb_{\rm TLS} $] TLS estimate of $\bb$ \eqref{eq:TLSest-k}

\item[$\bb_{\kappa} $] ECM parameter vector at the $\kappa^{\rm th}$ block \eqref{eq:KFprocess}
\item[$\hat \bb_{\kappa} $] RLS/TLS estimate of $\bb$ at time block $\kappa$ \eqref{eq:bk1hat}
\item[$\hat \bb_{\kappa|\kappa} $] TKF estimate of $\bb$ at time block $\kappa$ (Algorithm \ref{alg:RTLS-KF})

\item[$\bH$] Augmented information matrix \eqref{eq:augmentd_obs_cbatt_est}
\item[$\bH_\kappa $] Augmented information matrix at time block $\kappa$ \eqref{eq:augmentd_k}

\item[$i(k)$] True current through resistor $R$ \eqref{eq:noiseless}
\item[$i_c$]  A certain (constant) value for $i(k)$ \eqref{eq:SNR}
\item[$\bi$] Column vector of consecutive $i(k)$ \eqref{eq:obs_vec}

\item[$n_{\rm i}(k)$] Current measurement noise \eqref{eq:ni(k)}
\item[$n_{\rm v}(k)$] Voltage measurement noise \eqref{eq:ni(k)}
\item[$\bn_{\rm v}$] Voltage noise in the vector observation model \eqref{eq:obs_vec}
\item[$\bn$] Combined noise in the vector observation model \eqref{eq:vector_model_noisy}
\item[$\bn(\kappa)$] Noise in the $\kappa^{\rm th}$ measurement vector \eqref{eq:vk}

\item[$\bP_{\rm LS}$] Covariance of the LS estimator \eqref{eq:LScov}
\item[$\bP_{\rm TLS}$] Covariance of the TLS estimator \eqref{eq:TLScov}
\item[$\bP_\kappa$] Covariance of the RLS estimator at $\kappa^{\rm th}$ block \eqref{eq:InfoUpdate}
\item[$\bP_{\kappa|\kappa}$] Covariance of the TKF estimator at $\kappa$ (Algorithm \ref{alg:RTLS-KF})

\item[$R$] True resistance $R$ \eqref{eq:model0}
\item[$\hat R$] An estimate of $R$ \eqref{eq:noiseless}
\item[$\hat R_{\rm LS}$] Least square estimate of $R$ \eqref{eq:LSest}
\item[$\bar R_{\rm LS}$] Average of $\hat R_{\rm LS}$ \eqref{eq:R-LS-average}
\item[$\hat R_{\rm TLS}$]  Total least square estimate of $R$ \eqref{eq:TLSest}
\item[$\bR$] Information matrix related to $\bH$ \eqref{eq:info_matrix}
\item[$\bR_{\kappa}$] Information matrix related to $\bH_\kappa$ \eqref{eq:covariance_aumented_matrix}

\item[$\bS_{\kappa+1} $] Residual covariance of the RLS filter \eqref{eq:sp}
\item[$\sigma^2_v$] Variance of the voltage measurement noise \eqref{eq:volt-noise-cov}
\item[$\sigma^2_{\rm CRLB}$] CRLB in estimating $R$ \eqref{eq:crlb}
\item[$\sigma_{\rm LS}^2  $] Variance of the LS estimator \eqref{eq:varLS}
\item[$\bar \sigma_{\rm LS}$] Normalized SDE of the LS estimator \eqref{eq:NORMvar}

\item[$v(k)$] True voltage across resistor $R$ \eqref{eq:noiseless}
\item[$\bv_1$] First column of $\bV$ \eqref{eq:TLSest}
\item[$\bv_2$] Second column of $\bV$ \eqref{eq:TLSest}
\item[$\bV$] Eigenvector matrix of $\bR$ \eqref{eq:SVD}
\item[$\bV_\kappa$] Eigenvector matrix of $\bR_\kappa$ \eqref{eq:eigen_decomp}
\item[$\tilde \bv_\kappa$] Process noise in TKF \eqref{eq:KFprocess}

\item[$\bW_\kappa$] Filter gain at time $\kappa$ \eqref{eq:bk1hat}
\item[$\tilde \bw_\kappa$] Measurement noise in TKF \eqref{eq:KFmeas}

\item[$z_{\rm i}(k)$] Measured value of $i(k)$ \eqref{eq:ni(k)}
\item[$z_{\rm v}(k)$] Measured value of $v(k)$ \eqref{eq:model0}
\item[$\bz_{\rm i}$] Column vector of consecutive $z_{\rm i}(k)$ \eqref{eq:vector_model_noisy}
\item[$\bz_{\rm v}$] Column vector of consecutive $z_{\rm v}(k)$ \eqref{eq:vector_model_noisy}
\item[$\bz_{\rm v}(\kappa)$] $\kappa^{\rm th}$ observation vector of $\bz_{\rm v}$ \eqref{eq:vk}
\item[$\bz_\kappa $] Measurement at the $\kappa^{\rm th}$ time block to the TKF \eqref{eq:KFmeas}

\end{abbreviations}

\section{Least Square Estimate Under Ideal Conditions}
\label{sec:LS_ideal}

{The least square approach is used to create a model that best fits a given set of data by minimizing the sum of the squares of the error in the measurements.
This approach is widely used in many applications such as image processing \cite{dong2018effective} and signal modeling \cite{xu2017recursive} and equivalent circuit model parameter identification in batteries \cite{sun2019adaptive}.}
\subsection{Theory}
\label{subsec:LStheory}

\begin{figure}[h]
\begin{center}
\includegraphics[width=.5\columnwidth]{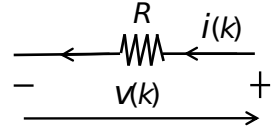}
\end{center}
\caption{
{\bf Estimating the value of a resistance.}
Current $i(k)$ is applied and the resulting voltage across the resistance {$v(k)$} is measured. 
}
\label{fig:model0}
\end{figure}

Figure \ref{fig:model0} shows a scenario \cite{sreenath2017resistive} where the unknown resistance $R$ needs to be estimated. 
By sending a current $i(k)$ through the resistor and measuring the resulting voltage $v(k)$ across the resistor, the resistance can be simply estimated to be 
\begin{align}
\hat R = v(k)/i(k) 
\label{eq:noiseless}
\end{align}
where $k$ indicates the sampling time instance. 
However, this is true only when the measured voltage and current are noise-free. 

First, let us consider a case where the voltage measurements are noisy (whereas the current through the resistor is assumed perfectly known).
Under this assumption, the measured voltage across the resistor is modelled as 
\begin{align}
z_{\rm v}(k) = i(k) R + n_{\rm v}(k)
\label{eq:model0}
 \end{align}
 where $z_{\rm v}(k) $ is the measured voltage at time $k,$
 $i(k)$ is the current at time $k,$
 and
 $n(k) $ is the voltage measurement noise which is assumed to be independently and identically distributed (i.i.d.) Gaussian with zero-mean and standard deviation $ \sigma_{\rm v} .$ 
 Considering $m$ consecutive observations, the above {\em observation model} can be written in vector form as
 \begin{align}
\bz_{\rm v} = \bi R + \bn_{\rm v}
\label{eq:obs_vec}
 \end{align}
 where $\bz_{\rm v}, \bi$ and $\bn_{\rm v}$ are column vectors of voltage, current and noise, respectively, of length $m$.
The covariance of the measurement noise vector, under the i.i.d. assumption, is given by 
\begin{align}
E\left\{  \bn_{\rm v}  \bn_{\rm v}^T \right\} = \sigma_{\rm v} ^2 \bI
\label{eq:volt-noise-cov}
\end{align}
where 
$\bI$ is an $m \times m$ identity matrix. 
The least squares estimator \cite{bar2004estimation} of the resistance is given as
\begin{align}
\hat R_{\rm LS} &= \frac{\bi^T \bz_{\rm v}} { \bi^T \bi } 
\label{eq:LSest}
\end{align}

It can be shown that the LS estimate \eqref{eq:LSest} is {\em unbiased}, i.e., 
\begin{align} 
 b_{\rm LS} &=  R -  E \left\{ \hat R_{\rm LS}  \right\}  \nonumber \\
&=  R -  \mu_{\rm LS}   = 0
\label{eq:muLS}
\end{align}
where 
$\mu_{\rm LS} $ denotes the expected value of $\hat R_{\rm LS}$ and
$ b_{\rm LS} $ denotes the bias of the least square estimator. 
Further, it can also be shown that the LS estimator in \eqref{eq:LSest} is {\em efficient}, i.e.,
\begin{align}
\sigma_{\rm LS}^2  =  E \left\{ \left( \hat R_{\rm LS} - \mu_{\rm LS} \right)^2  \right\} =    \sigma_{\rm CRLB}^2
\label{eq:varLS}
\end{align}
where $ \sigma_{\rm CRLB}^2$ is the Cramer-Rao lower bound, a theoretical lower bound on the variance $\sigma_{\rm LS}^2 $ based on the model described {\em regardless of the estimation approach.}
It can be shown that 
\begin{align}
 \sigma_{\rm CRLB}^2 & = \frac{  \sigma_{\rm v} ^2 } { \bi^T \bi } = \sigma_{\rm v} ^2 \left({\sum_{k=1}^{m}} i(k)^2 \right)^{-1}
 \label{eq:crlb}
\end{align}

From \eqref{eq:crlb}, one can conclude the following about the variance of the least square estimator:
\begin{enumerate}
\item[(a)] {\em Measurement error effect:} 
The variance of the LS estimator is proportional to $\sigma_{\rm v} ^2$, i.e., when the measurement noise increases, so does the {theoretical lower bound of the estimation variance}. 
\item[(b)] {\em Effect of number of samples: } {As the number of observations $m$ increases, the theoretical lower bound variance of the estimation decreases}.
\item[(c)] {\em Signal strength effect: }
The {theoretical lower bound }variance of the estimation error gets smaller when the strength of the excitation, i.e., the absolute value of $i(k)$, becomes larger. 
\end{enumerate}
In summary, the CRLB in \eqref{eq:crlb} helps us to arrive at the conclusion that lower measurement noise, higher number of samples and larger values of the current $i(k)$ are preferred for better estimation of $R$. 

In the next subsection we will demonstrate the above using Monte Carlo simulations.
This demonstration will later serve as a comparison when noise is introduced into the current measurement, i.e., $i(k)$, and the focus of this paper.

\subsection{Monte-Carlo Demonstration}
\label{subsec:LSresults}

In this section, the goal is to estimate the bias and variance of the least square estimator in \eqref{eq:LSest} and to demonstrate their expected behaviors under the model assumptions. 
In order to do that, we simulate the measured voltage $z_{\rm v}(k)$ in \eqref{eq:model0} based on the following assumptions:
\begin{itemize} \setlength\itemsep{0pt}\setlength\parskip{0pt}
 \item
 Current is assumed to be known and constant $i(k) = i_c = 2 \, {\rm A}.$ 
 \item
  Resistance is set to be a constant $R= 0.25 \, \Omega .$
  \item
  Measurement noise $n_{\rm v}(k)$ is zero-mean white Gaussian with standard deviation (s.d.) $ \sigma_{\rm v}  = 1 \, {\rm mV} .$
  \item
 $m= 100$ (length of the measurement vector $\bz_{\rm v}$).
 \end{itemize}
 Based on the above assumptions, the vector observation model \eqref{eq:obs_vec} was used to simulate the noisy voltage observations; then, the LS estimator \eqref{eq:LSest} was applied and an estimate $\hat R_{\rm LS}$ is obtained.  

The above simulation was repeated for $j=1, \ldots, {M}=1000$ Monte-Carlo runs.
Each time, the computed LS estimate is denoted as $\hat R_{\rm LS}(j), \,\, j=1, \ldots, M$. 
Now, the {\em normalized bias} and {\em normalized standard deviation of error (SDE)} of the LS estimator are computed as 
\begin{align}
\bar b_{\rm LS} & = \left( \frac{ \bar{ R}_{\rm LS} - R}{R} \right) \times 100 \label{eq:NORMbias} \\
\bar \sigma_{\rm LS} &=  \frac{1}{R}  \sqrt{ \frac{1}{M} \sum_{j=1}^{M} \left( \hat R_{\rm LS}(j) -  R  \right)^2} \times 100
\label{eq:NORMvar}
\end{align}
where 
\begin{align}
 \bar{ R}_{\rm LS} =\frac{1}{M} \sum_{j=1}^{M}\hat R_{\rm LS}(j) 
 \label{eq:R-LS-average}
\end{align}
The bias and variance are defined in normalized form in order for them to be meaningful regardless of the application, e.g., in some applications, such as smart phones, typical value of $i(k)$ is in mere Amperes whereas in other applications, such as in electric vehicles, typical value of $i(k)$ is in tens and hundreds of Amperes. 

The Monte-Carlo experiment detailed so far is designed for a voltage measurement noise standard deviation of $ \sigma_{\rm v}  = 1 \, {\rm mV} $. 
It is important to note that measurement systems have varying degree of accuracy. 
Further, as it is discussed in Section \ref{subsec:LStheory}, the performance of an LS estimator is sensitive to the amplitude of the current through the resistor --- the higher the current $i(k)=i_c,$ the more accurate the LS estimate. 
In order to understand the performance against varying current and noise levels the following {\em signal-to-noise ratio} \cite{jenq1996discrete} is defined
\begin{align}
{\rm SNR} = 20 \log_{10 } \left( \frac{i_c R}{ \sigma_{\rm v} } \right) 
\label{eq:SNR}
\end{align}
and the experiment is repeated for different values of SNR values shown in Table \ref{table:SNR} by changing the standard deviation of the noise. 
\begin{table}[h]
\caption{SNR and corresponding noise s.d. for a constant current of $i_c = 2 \,{\rm A}$}
\label{table:SNR}
\begin{center}
\begin{tabular}{|c|c|c|c|c|c|c|c|}
\hline
SNR (dB) & $0$   &  $10$   &  $20$   &  $30$   &  $40$   &  $50$  \\
\hline
$ \sigma_{\rm v}  ({\rm mV})$ & 2000 & 633 & 200 & 63 & 20 & 6 \\
\hline
\end{tabular}
\end{center}
\end{table}

\begin{figure}
\begin{center}
\includegraphics[width=.9\columnwidth]{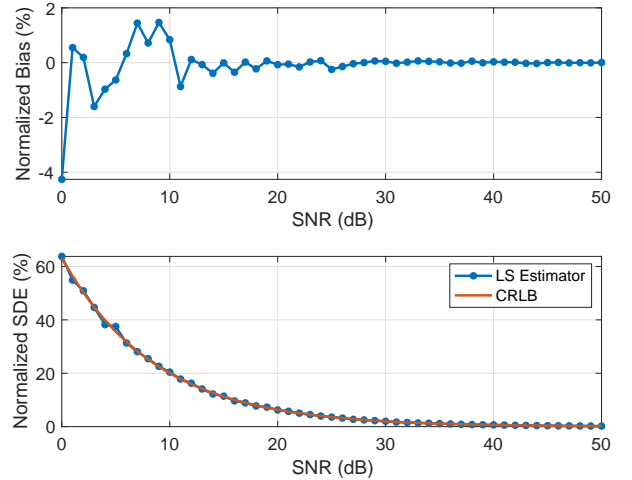}
\end{center}
\caption{
{\bf Least square estimator in ideal conditions. }
When the current is known, the LS estimate of $R$ is both {\em unbiased} and {\em efficient}.
The bias and variance plots are the average of 1000 Monte Carlo runs. 
}
\label{fig:ideal_case}
\end{figure}

Figure \ref{fig:ideal_case} shows the results of the Monte-Carlo experiment detailed so far in this section for the SNR values shown in Table \ref{table:SNR}. 
The bias and variance plots shown in Figure \ref{fig:ideal_case} are averaged over $M=1000$ Monte-Carlo runs. 
The top plot in Figure \ref{fig:ideal_case} shows the bias of the LS estimator that is computed according to \eqref{eq:NORMbias}.  
It can be observed that averaged bias is zero mean.

The second plot in Figure \ref{fig:ideal_case} shows the variance of the LS estimator that is computed according to \eqref{eq:NORMvar} and the CRLB at each SNR that was computed based on \eqref{eq:crlb}. 
The CRLB and the variance of the estimator are shown to be identical at all SNR levels -- this indicates that the LS estimator is efficient regardless of the SNR. 

\section{Noisy Model and Total Least Squares Estimation} 
\label{sec:NoisyModel}

In some applications it is required to estimate the resistance of a system while it is being used, i.e., in {\em real-time}. 
For example, the resistance of a battery changes with the temperature and it needs to be estimated in real-time so that its remaining power can be accurately estimated.  
When the battery is in use, the amount of current is determined by the end user, i.e., it needs to be measured so that the necessary parameters, such as its internal resistance, can be estimated. 
When both voltage and current are {\em measured}, the assumptions made in Section \ref{sec:LS_ideal} are not valid anymore and the LS estimate becomes biased and inefficient --- we demonstrate this later in Subsection \ref{sec:MCdemo_LSvsTLS}. 
In this section, we propose an alternate solution, based on the total least squares (TLS) approach. 
\blue{include TLS references.}

\subsection{Theory}

Consider the case where the current through the resistor also needs to be measured. 
In this case, the observation model can be written as
\begin{align}
z_{\rm v}(k) = z_{\rm i}(k) R + n(k)
\label{eq:noisymodel}
\end{align}
where 
\begin{align}
z_{\rm v}(k) &= v(k) + n_{\rm v} (k) \label{eq:nv(k)} \\
z_{\rm i}(k)  &= i(k) + n_{\rm i}(k) \label{eq:ni(k)}
\end{align}
are the measured voltage and current, respectively;
$n_{\rm i} (k)$ is the current measurement noise that is assumed to be i.i.d. zero mean white with standard deviation $\sigma_i;$ as before, 
$n_{\rm v} (k)$ is the voltage measurement noise that is assumed to be i.i.d. zero mean white with standard deviation $\sigma_{\rm v} .$

Similar to \eqref{eq:obs_vec}, considering $m$ measurements, the vector observation model can be written as 
 \begin{align}
\bz_{\rm v} = \bz_{\rm i} R + \bn
\label{eq:vector_model_noisy}
 \end{align}
where $\bz_{\rm v} $, $\bz_{\rm i} $ and $\bn$ are vectors of length $m.$
The least square estimator, reviewed in Section \ref{sec:LS_ideal}, assumes that the model (in this case $ \bz_{\rm i} $ is the model) is perfectly known. 
When the model is not perfectly known the LS estimator cannot be guaranteed to be unbiased and efficient.

The TLS method is able to model uncertainty in both the measurement $\bz_{\rm v}$ and the model $\bz_{\rm i}$ itself; 
For more details about the TLS method, the reader is referred to \cite{markovsky2007overview} and the references therein.
Also, the TLS approach was applied to battery capacity estimation in \cite{bfg_capest_2014}. 

Let us construct the following {\em augmented observation matrix}
\begin{eqnarray}
\bH = \left[ \bz_{\rm v} \,\,\,\, \bz_{\rm i} \right] \label{eq:augmentd_obs_cbatt_est}
\end{eqnarray}
The information matrix associated with the augmented observation matrix is defined as
\begin{eqnarray}
\bR = \bH^T \bH
\label{eq:info_matrix}
\end{eqnarray}
where $(\cdot)^T$ denotes transpose.
The eigendecomposition of $\bR$ can be written in the following format
\begin{eqnarray}
\bR  = \bV  \bLambda  {\bV}^T 
\label{eq:SVD}
\end{eqnarray}
where  $\bLambda $ is a $2 \times 2$ diagonal matrix of nonnegative (and real) eigenvalues arranged from the largest to the smallest, i.e., $ \bLambda (1,1)$ denotes the largest eigenvalue and $ \Lambda_\kappa(2,2)$ denotes the smallest.
Each column of the $2 \times 2$ matrix $\bV= \left[ \bv_1, \,\, \bv_2 \right]$ has the corresponding eigenvectors, i.e., 
 $\bv_1$ is the eigenvector corresponding to $ \bLambda (1,1)$ and $\bv_2$ is the eigenvector corresponding to $ \bLambda(2,2)$. 

Now, the TLS estimate of $R$ is given by \cite{van1991total}
\begin{align}
\hat R_{\rm TLS} &= - \frac{ \bv_2(1) }{\bv_2(2)}
\label{eq:TLSest}
\end{align}

In the next subsection, we compare the performances of the LS and TLS estimators using Monte-Carlo simulations.

\subsection{Monte-Carlo Demonstration}
\label{sec:MCdemo_LSvsTLS}

The goal in this section is to compare the performance of the LS and TLS estimators when there is a model uncertainty, i.e., when both voltage $v(k)$ and current $i(k)$ are measured as shown in \eqref{eq:nv(k)} and  \eqref{eq:ni(k)}, respectively. 
We will use similar simulation parameters to the ones used in Section \ref{subsec:LSresults} with the following highlights/modifications:
\begin{itemize}
\item The true current is kept constant at $i(k) = i_c = 2 \,\, {\rm A}$ and the true resistance is also kept constant at $R = 0.25 \,\,  \Omega.$
\item
With the above assumption the definition of the SNR is the same as in \eqref{eq:SNR}.
\item
Using the above parameters, the voltage and current measurements are generated for different SNR values ranging from -30 dB to 50 dB according to the model defined in \eqref{eq:noisymodel}--\eqref{eq:ni(k)}.
\item
Based on the simulated voltage and current measurements the following two estimates are obtained: 
\begin{itemize}
\item {\em LS estimate:} The LS estimate $\hat R_{\rm LS}$ is computed using \eqref{eq:LSest} by assuming that the model is $\bz_{\rm i}$
\item {\em TLS estimate:} The TLS estimate is computed according to \eqref{eq:TLSest}
\end{itemize}
\end{itemize}

\begin{figure}
\begin{center}
\includegraphics[width=.9\columnwidth]{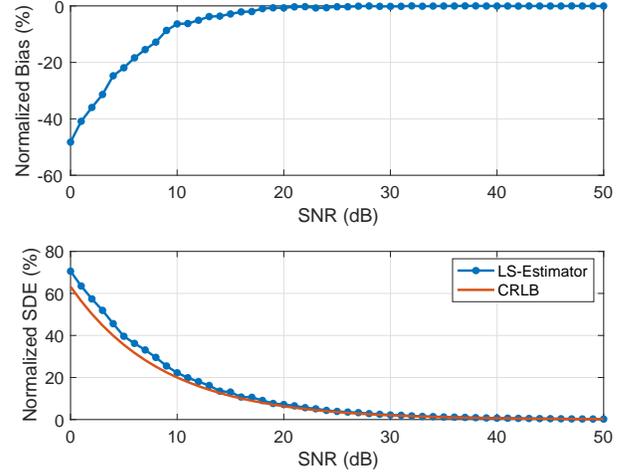}
\end{center}
\caption{
{\bf LS estimate of a noisy model. }
When there is noise in the model, as shown in \eqref{eq:noisymodel}, the LS estimator is neither unbiased and nor is it efficient. 
The top plot shows that the bias of the estimator is incremental with decreasing SNR where as the variance deviates from the theoretical bound (which was defined for perfectly known model \eqref{eq:obs_vec}).
}
\label{fig:mistmatch_demo_1}
\end{figure}

Figure \ref{fig:mistmatch_demo_1} shows the normalized bias and normalized standard deviation of the LS estimator, defined in \eqref{eq:NORMbias} and \eqref{eq:NORMvar}, respectively. 
The observation here is that with the model uncertainty, the LS estimate becomes biased --- especially when the SNR is low. 
Further, the lower plot in Figure \ref{fig:mistmatch_demo_1} shows that the variance of the estimator is not meaningful anymore.
{It can be noted that the normalized standard deviation, defined in \eqref{eq:NORMvar}, is based on biased estimates and renders itself meaningless.
In other words, a biased estimate cannot be efficient \cite{pugachev2014probability}, which is the case for the least square algorithm when the SNR is low.} 
Due to this reason, we will define the estimation error variance in terms of the true parameters in future demonstrations. 

\begin{figure}
\begin{center}
\includegraphics[width=.9\columnwidth]{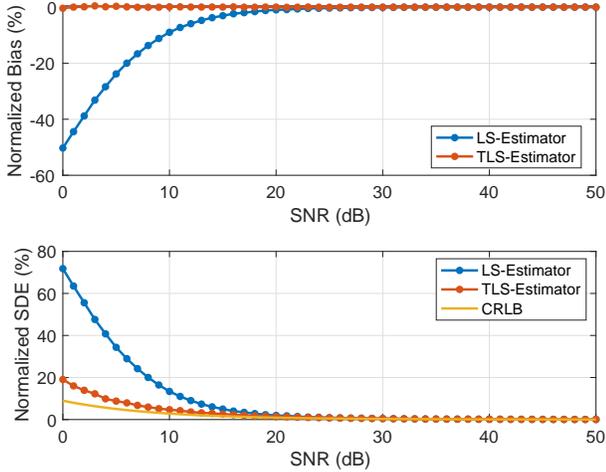}
\end{center}
\caption{
{\bf Comparison of LS estimate with TLS estimate. }
Both the LS estimator and the TLS estimator used $m=500$ observations. 
The bias and MSE showed in this figure were averaged over 1000 Monte Carlo runs. 
}
\label{fig:TLS_demo1}
\end{figure}

Figure \ref{fig:TLS_demo1} shows a comparison of the LS estimate and TLS estimate on the same plot. 
The top plot in Figure \ref{fig:TLS_demo1} shows that as the LS estimate becomes biased at low SNR values, the TLS estimate remains nearly unbiased. 
Further, the variance of the TLS estimate remains meaningful for all SNR values -- the conclusion here is that at low SNR values, the TLS estimator outperforms the LS estimator. 
When the SNR is high, both the LS and TLS estimators yield unbiased and efficient estimates. 
This raises an important observation that the parameter estimator should be decided based on the practical value of SNR in a particular application.  
If it is known that the SNR will be high, then the LS estimator should be selected because of its computational advantage over the TLS estimate. 
However, we will point out later in Section \ref{sec:time-var-snr} that most practical applications are subjected to time-varying SNR, where the SNR is likely to drop to very low values. 

\begin{figure}
\begin{center}
\includegraphics[width=.9\columnwidth]{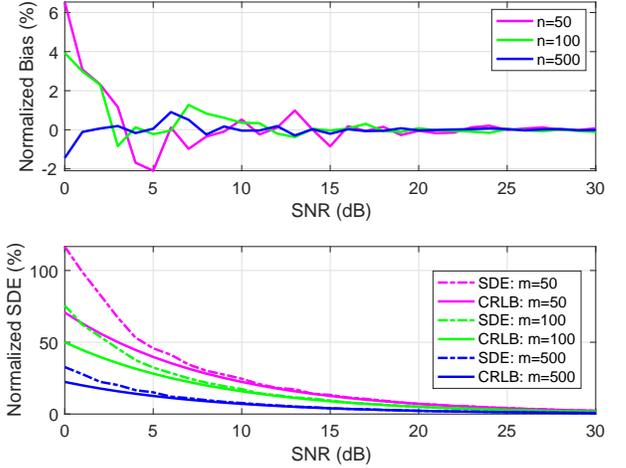}
\end{center}
\caption{
{\bf Performance of TLS with number of samples. }
The bias of the TLS estimate remains zero at all SNR values whereas the variance reduces and approaches the CRLB with increasing number of observations in a batch. 
}
\label{fig:TLS_demo2}
\end{figure}

Figure \ref{fig:TLS_demo2} shows the performance of TLS with varying number of samples in terms of bias (top) and standard deviation of error (bottom). 
It must be noted that as the number of observations increases, the CRLB decreases.
To further illustrate this, let us simplify the CRLB (see \eqref{eq:crlb}) to the present assumption, i.e., $i(k) = i_c$:
\begin{align}
 \sigma_{\rm CRLB}^2 & = \left(\frac{  \sigma_{\rm v} ^2 } {i_c^2 } \right) \left( \frac{1}{m} \right) 
\end{align}
where $m$ is number of observations. 
That is, the estimation error variance is inversely proportional to the number of observations. 

Another observation from Figure \ref{fig:TLS_demo2} is that, for a given SNR, increase in the number of samples $m$ helps to move the estimator towards efficiency. 
However, increasing the number of samples may not always be possible due to computational bottlenecks and other factors, such as the possibility of changes in battery parameters due to environmental factors. 
In the next section, we will discuss recursive estimation approaches that are designed to retain the best of both worlds:
If the parameter remains constant, the performance of the recursive filter will approach its theoretical limit. 
If the parameter changes, the recursive filter will ``track'' it in the best possible manner via fading memory parameter. 

\section{Recursive Estimation}
\label{sec:RecusiveEST}

{
Although LS and TLS estimates are good ways to model a system, their drawback is that they are expensive to run in real time because when there is new measurement it can either drop an old measurement and replace it with the new one which can decrease the accuracy of the algorithm or increase the number of measurements which makes the algorithm more complex over time and unfeasible.
In order to solve this problem two alternative algorithms are used recursive least square (RLS) and recursive total least square (RTLS).
RLS is widely adopted for battery identification, a RLS algorithm is proposed in \cite{lao2018novel,sun2019adaptive} where a variable forgetting factor that is changed based on a factor that is defined as the difference between the predicted measurements and the measurement itself.
The authors in \cite{xia2018online} proposed the use of multiple forgetting factors when there are multiple parameters are estimated, the appropriate forgetting factor is set for each individual parameter.
In \cite{wei2017online} a similar RTLS approach is proposed to identify battery parameters and estimate the state of charge.}

Let us write the observation model \eqref{eq:vector_model_noisy} with a slightly different, but more generic, notation as follows
\begin{eqnarray}
\bz_{\rm v}(\kappa)=   \bA_\kappa \bb + \bn(\kappa) \label{eq:vk}
\end{eqnarray}
where 
$\kappa$ denotes the batch number,
$\bz_{\rm v}(\kappa)$ is the $m\times 1$ vector of observations, 
$\bA_\kappa$ is the $m\times n$ model, 
$\bb$ is the $n\times 1$ parameter vector to be estimated, 
and 
$\bn_\kappa$ is the $m\times 1$ noise vector which is assumed zero-mean i.i.d. Gaussian with its  $m\times m$ covariance matrix written as
\begin{equation}
\bSigma = E\left\{  \bn(\kappa)  \bn(\kappa)^T  \right\}
\label{eq:SIGMA}
\end{equation}
For the case of \eqref{eq:vector_model_noisy}, the above model simplifies as follows:
\begin{align}
\bz_{\rm v}(\kappa) = \bz_{\rm v}, \,\, \bA_\kappa = \bz_{\rm i},  \,\,  \bb = R,  \,\, \bn(\kappa) = \bn, \,\,  \bSigma = \sigma_{\rm v} ^2  \bI_m
\label{eq:assumptions}
\end{align}
where $\bI_m$ is an $m \times m$ identity matrix. 
It is worth to emphasize that the parameter $\bb$ has no subscript $\kappa$ which implies that the parameter is assumed to be a constant. 
The (recursive) estimation schemes and the performance assessment method that follow all will assume the unknown parameter to be a constant.

In the remainder of this section, we will briefly summarize the recursive least squares (RLS) filter and some alternative recursive filters proposed for the recursive estimation of $\bb$.

\subsection{Recursive Least Squares Filter}

Given the $\kappa^{\rm th}$ batch of observation $\bz_{\rm v}(\kappa)$ along with the model $\bA_\kappa$, 
the LS estimator of the parameter $\bb$ is given by
\begin{align}
\hat \bb_{\rm LS} = \left( \bA_\kappa^T\bSigma^{-1} {\bA_\kappa} \right)^{-1}  {\bA_\kappa}^T\bSigma^{-1} \bz_{\rm v}(\kappa) \label{eq:bkhat}
\end{align}
which, for the assumptions in \eqref{eq:assumptions} will reduce to \eqref{eq:LSest}. 
The corresponding LS estimation error covariance is given as 
\begin{align}
\bP_{\rm LS} = \left( \bA_\kappa^T\bSigma^{-1} {\bA_\kappa} \right)^{-1} 
\label{eq:LScov}
\end{align}

Next, we summarize the recursive least squares (RLS) filter for the estimation of $\bb$. 
Let us assume that the estimates corresponding the the $\kappa^{\rm th}$ batch is $\hat \bb_\kappa  $ and the corresponding estimation error covariance is $\bP_\kappa$. 
When the next batch of observation $\bv_{\kappa+1}$ becomes available, the RLS filter \cite{bar2004estimation} updates the parameter estimate as follows:
\begin{eqnarray}
\hat \bb_{\kappa+1}= \hat \bb_{\kappa} + \bW_{\kappa + 1} \left( \bv_{\kappa+1} - \bA_{\kappa+1} \hat \bb_{\kappa} \right) \label{eq:bk1hat}
\end{eqnarray}
where 
\begin{eqnarray}
\bW_{\kappa+1} &= \bP_{\kappa+1}\bA_{\kappa+1}^T \bS_{\kappa+1}^{-1} \label{eq:updategain}
\end{eqnarray}
and
\begin{align}
\bP^{-1}_{\kappa+1}&= \bP^{-1}_\kappa + \bA^T_{\kappa+1}\bSigma^{-1} \bA_{\kappa+1} \label{eq:InfoUpdate} \\
\bS_{\kappa+1} &=\bA_{\kappa+1}\bP_{\kappa+1}\bA^T_{\kappa+1} + \bSigma \label{eq:sp}
\end{align}
where $\bP_\kappa$ is the estimation error covariance of the RLS filter at the time block $\kappa.$

Algorithm \ref{alg:RLS} summarizes one recursive step of the RLS filter that requires the following variables as input:
RLS estimate from the previous batch $\hat \bb_\kappa,$
information (inverse of the RLS estimation error covariance)  from the previous batch, $\bP_{\kappa}^{-1},$
new measurement vector $\bv_{\kappa+1}$, 
and 
the new observation matrix (or model) $\bA_{\kappa+1}$. 
The outputs of the RLS filter are the 
updated estimate, $\hat \bb_{\kappa+1},$ and the updated information matrix (i.e., inverse of the estimation error covariance), $\bP_{\kappa+1}^{-1}.$
\begin{algorithm}[H]
\caption{\\
$
{\big[\hat \bb_{\kappa+1},  \bP^{-1}_{\kappa+1} \big] }  
=\rm{RLS}
\small{ [\hat \bb_\kappa ,\bP^{-1}_\kappa, \bv_{\kappa+1},\bA_{\kappa+1}]}
$
}
\label{alg:RLS}
\begin{algorithmic}[1]
\State{Update information: $\bP^{-1}_{\kappa+1}=\bP^{-1}_{\kappa}+\bA^T_{\kappa+1}\Sigma^{-1}\bA_{\kappa+1} $}
\State{Update residual cov.: $\bS_{\kappa+1} =\bA_{\kappa+1}\bP_{\kappa+1}\bA^T_{\kappa+1} + \bSigma$ }
\State{Update gain: $\bW_{\kappa+1} = \bP_{\kappa+1}\bA_{\kappa+1}^T \bS_{\kappa+1}^{-1}$ }
\State{Update parameter: $\hat \bb_{\kappa+1}= \hat \bb_{\kappa} + \bW_{\kappa + 1}( \bv_{\kappa+1} - \bA_{\kappa+1} \hat \bb_{\kappa})$}
\end{algorithmic}
\end{algorithm}

In the next subsection, we briefly summarize the total least squares filter \cite{van1991total} and present an approach to implement it in a recursive manner.

\subsection{Recursive Total Least Squares Filter}

For the $\kappa^{\rm th}$ observation block, the $m \times (n+1)$ augmented observation matrix is given as 
\begin{eqnarray}
\bH_\kappa = \left[ \bz_{\rm v}(\kappa) \,\,\,\, \bA_\kappa \right] \label{eq:augmentd_k}
\end{eqnarray}
The information matrix associated with the augmented observation matrix is defined as
\begin{eqnarray}
\bR_{\kappa} = \bH_\kappa^T \bH_\kappa \label{eq:covariance_aumented_matrix}
\end{eqnarray}
Now, the eigendecomposition of $\bR_{\kappa} $ can be written in the following format
\begin{eqnarray}
\bR_{\kappa}  = \bV_\kappa  \bLambda_\kappa  { \bV_\kappa }^T  \label{eq:eigen_decomp}
\end{eqnarray}
where  $\bLambda_\kappa $ is an $(n+1) \times (n+1)$ diagonal matrix of nonnegative (and real) eigenvalues arranged from the largest to the smallest, i.e., $ \bLambda_\kappa (1,1)$ denotes the largest eigenvalue and $ \Lambda_\kappa(n+1,n+1)$  denotes the smallest eigenvalue.
Each column of the $(n+1) \times (n+1)$ matrix $\bv(\kappa) = \big[ \bv_1, \,\, \bv_2 , \ldots, \bv_{n+1}\big]$ has the corresponding eigenvectors, i.e., 
the first column $\bv_1$ is the eigenvector corresponding to $ \bLambda_\kappa (1,1)$,
the second column $\bv_2$ is the eigenvector corresponding to $ \bLambda_\kappa (2,2)$, and so on. 

Let us write $ \bV_\kappa $ in the following partitioned format
\begin{align}
 \bV_\kappa  = 
\left[
\begin{array}{c|c}
\bV_{11} & \bV_{12} \\
\hline 
\bV_{21} & \bV_{22} \\
\end{array}
\right]
\end{align}
where 
$\bV_{11} $ is an $n \times n$ matrix,
$\bV_{12} $ is an $n \times 1$ vector,
$\bV_{21} $ is an $1 \times n$ vector,
and
$\bV_{22} $ is a scalar. 
Now, the TLS estimate of $\bb$ is given as \cite{van1991total}
\begin{align}
\hat \bb_{\rm TLS} = - \bV_{12} \bV_{22}^{-1}
\label{eq:TLSest-k}
\end{align}
There are various derivations for the covariance of the TLS estimate --- none of them are exact;
one of the most adopted ones is \cite{van1991total}
\begin{align}
\bP_{\rm TLS} = \left( \bA_\kappa^T \bA_\kappa - \sigma_{n+1}^2 \bI \right)^{-1}
\label{eq:TLScov}
\end{align}

Similar to the RLS filter, the TLS estimate above can be obtained in a recursive form. 
For this, the information matrix in (\ref{eq:covariance_aumented_matrix}) is updated with a fading memory, as follows
\begin{eqnarray}
\bR_{\kappa+1} = \lambda \bR_{\kappa} +  \frac{(\bH_{\kappa+1})^T \bH_{\kappa+1} }{m - 1} \label{eq:tls_cov_update}
\end{eqnarray}
and the TLS estimate is obtained each time using \eqref{eq:TLSest-k}. 

Algorithm \ref{alg:RTLS} summarizes one recursive step of the recursive TLS filter that requires the following variables as input:
information matrix from the previous batch, $\bR_{\kappa},$
new measurement vector $\bv_{\kappa+1}$, 
and 
the new observation matrix (or model) $\bA_{\kappa+1}$. 
The output of the TLS filter are the 
present estimate, $\hat \bb_{\kappa+1},$
updated information matrix, $\bR_{\kappa+1}$
and the covariance of the present TLS estimate, $\bP_{\kappa+1}$.

\begin{algorithm}[H]
\caption{\\
$
{\left[\hat \bb_{\kappa+1}, \bR_{\kappa+1}, \bP_{\kappa+1} \right] }  
=\rm{TLS}
\small{ \left[\bR_\kappa, \bv_{\kappa+1}, \bA_{\kappa+1}  \right]}
$
}
\label{alg:RTLS}
\begin{algorithmic}[1]
\State{Construct augmented observation matrix:}
\Statex{$\bH_{\kappa+1} = \left[ \bv_{\kappa+1} \,\,\,\, \bA_{\kappa+1} \right] $}
\State{Update the information matrix:}
\Statex{$\bR_{\kappa+1} = \lambda \bR_{\kappa} +  \left( {(\bH_{\kappa+1})^T \bH_{\kappa+1} } \right) / {(m - 1)} $}
\State{Perform EVD: ${\bR}_\kappa   = \bV_\kappa  \bLambda_\kappa   {\bV_\kappa }^T   $}
\State{Compute TLS estimate: 
$ \hat \bb_{\kappa+1} = - \bV_{12} \bV_{22}^{-1} $}
\State{Compute TLS covariance: 
$\bP_{\kappa+1} = \left( \bA_{\kappa+1}^T \bA_{\kappa+1} - \sigma_{n+1}^2 \bI \right)^{-1}$
}
\end{algorithmic}
\end{algorithm}

It will later be elaborated in Subsection \ref{subsec:RecMCdemo} that the performance of the TLS algorithm, summarized in Algorithm \ref{alg:RTLS}, has a trade-off between higher estimation error variance and higher convergence time depending on the forgetting factor $\lambda$. 
In the next subsection, we propose an approach in order to improve the performance of the TLS filter.

\subsection{The Total Kalman Filter}

In this subsection, we assume that the true parameter $\bb$ might undergo small perturbations over time. 
Let us mode this process through the following {\em process model}
\begin{align}
\bb_{\kappa+1} &= \bb_{\kappa} + \tilde \bv_\kappa
\label{eq:KFprocess}
\end{align}
where $\tilde \bv_\kappa $ is a zero-mean Gaussian white noise process that is assumed to have zero mean and covariation given by
\begin{align}
E\{ \tilde \bv_\kappa  \tilde \bv_\kappa^T  \} = \bQ = \gamma \bI_n
\label{eq:gamma}
\end{align}
and $\gamma$ is a small number. 

Let us assume that the parameter is in the presence of noise, i.e., the {\em observation model} is written as 
\begin{align}
\bz_\kappa & = \bb_\kappa + \tilde \bw_\kappa
\label{eq:KFmeas}
\end{align}
where $\tilde \bw_\kappa$ is assumed to be zero-mean white Gaussian noise.

With \eqref{eq:KFprocess} and \eqref{eq:KFmeas} as process and measurement models, respectively, a Kalman filter is used to obtain the minimum mean square estimates of the parameter $\bb$ based on the observations $\bz_\kappa = \hat \bb_{\kappa}$ from Algorithm \ref{alg:RTLS}. 
Since the TLS estimates were shown to be unbiased in Section \ref{sec:NoisyModel} we may assume that the measurement noise $\bw_\kappa$ is zero mean.
Also, the covariance of the TLS estimates, i.e., the covariance of $ \bw_\kappa$, is given by \eqref{eq:TLScov}. 

In summary, the proposed total Kalman filter (TKF) algorithm uses the TLS estimates from Algorithm \ref{alg:RTLS}  as the measurement in the linear-Gaussian state-space model given in \eqref{eq:KFprocess} and \eqref{eq:KFmeas}.
Algorithm \ref{alg:RTLS-KF} summarizes one recursive step of the TKF that requires the following variables as input:
TKF estimate from the previous batch $\hat \bb_{\kappa|\kappa} ,$
corresponding estimation error covariance from the previous batch, $\bP_{\kappa|\kappa},$
information matrix from the previous batch, $\bS_{\kappa},$
new measurement vector $\bv_{\kappa+1}$, 
and 
the new observation matrix (or model) $\bA_{\kappa+1}$. 
The output of the TKF are the 
present estimate, 
$\hat \bb_{\kappa+1| \kappa+1},$ 
corresponding estimation error covariance, $\bP_{\kappa+1|\kappa+1},$
and the corresponding updated information matrix, $\bS_{\kappa+1}.$

\begin{algorithm}[h]
\begin{footnotesize}
\caption{\\
$[ \hat\bb_{\kappa+1|\kappa+1},\bP_{\kappa+1|\kappa+1} , \bS_{\kappa+1}]$  = \\
{\rm TKF}$ { \left( \hat\bb_{\kappa|\kappa},\bP_{\kappa|\kappa}, \bS_\kappa,  \bA_{\kappa+1} , \bv_{\kappa+1}\right)}$
}
\label{alg:RTLS-KF}
\end{footnotesize}
\begin{algorithmic}[1]
\State{State prediction: $ \hat \bb_{\kappa+1|\kappa}=\hat \bb_{\kappa|\kappa} $}
\State{State prediction cov.: $ \bP_{\kappa+1|\kappa}= \bP_{\kappa|\kappa}+\bQ $ }
\State{Measurement prediction: $\hat \bv_{\kappa+1|\kappa}=  \hat \bb_{\kappa+1|\kappa}$ }
\State{TLS estimates: $
{\left[\hat \bb_{\kappa+1}, \bS_{\kappa+1}, \bSigma \right] }  
=\rm{TLS}
\small{ \left[\bS_\kappa, \bv_{\kappa+1}, \bA_{\kappa+1}  \right]}
$}
\State{Measurement residual: $ \bnu_{\kappa+1}=  \hat \bb_{\kappa+1}-\hat \bv_{\kappa+1|\kappa} $ }
\State{Innovation cov.: $ \bS_{k+1}= \bSigma + \bP_{\kappa+1|\kappa} $}
\State{Filter gain: $ \bW_{\kappa+1}=  \bP_{\kappa+1|\kappa}  \bS_{k+1}^{-1}$}
\State{State est.: $ \hat \bb_{\kappa+1|\kappa+1}= \hat \bb_{\kappa+1|\kappa}+ \bW_{\kappa+1} \bnu_{\kappa+1}  $}
\State{State est. cov.:  
$\bP_{\kappa+1|\kappa+1}=  \bP_{\kappa+1|\kappa}- \bW_{\kappa+1}\bS_{\kappa+1}\bW_{\kappa+1}^T  $ }
\end{algorithmic}
\end{algorithm}

\subsection{Posterior CRLB}

Earlier in Section \ref{sec:LS_ideal}, we saw that the CRLB is a theoretical lower bound on the estimation error covariance given a batch of $m$ observations. 
When more observations become available, in the form of a batch of $m$ observations at a time, the CRLB will continue to drop. 
Under the assumptions made in \eqref{eq:assumptions} and assuming that the current is constant ($i[k]=i_c$) {it} reduces to the following scalar
\begin{align}
 \sigma_{\rm CRLB}^2 (\kappa)& = \left(\frac{  \sigma_{\rm v} ^2 } {i_c^2 } \right) \left( \frac{1}{m \kappa} \right) 
 \label{eq:crlb-kappa}
\end{align}
where $m$ is the length of the voltage measurement vector and $\kappa$ is the batch number.
It can be noticed that the CRLB is a function of $\kappa$ and that CRLB decreases with the arrival of new batches of measurements.

It must be noted that the Kalman filter is a Bayesian estimator, i.e., it can incorporate prior information in its estimation. 
For Kalman filter, the prior information is incorporated in the form of an initial distribution of the parameter, which is Gaussian by definition,  with mean $\hat \bb_{0}$ and covariance $\bP_{0|0}$.
When the prior information is good enough, the covariance of the Kalman filter could be even lower than the CRLB that was computed without accounting for the prior information.
That is, the CRLB shown in \eqref{eq:crlb-kappa} is not a lower bound anymore when Kalman filter is employed -- it could be lower than that. 
This new bound is known as the posterior CRLB (PCRLB).

For linear-Gaussian model under no model uncertainty, the PCRLB can be computed simply through the KF filter recursion as we summarized in Algorithm \eqref{alg:pcrlb}. 
Here, we assume that $\gamma$ in \eqref{eq:gamma} is negligible and that the observation matrix $\bA_{\kappa+1}$ in \eqref{eq:vk} is noiseless.

The input to the Algorithm \eqref{alg:pcrlb} are 
the computed PCRLB at the previous batch $\bP_\kappa,$
the {\em noiseless} model $\bA_{\kappa+1}$ in \eqref{eq:vk},
and the noise covariance of the observation model $\bSigma$, shown in \eqref{eq:SIGMA}. 
It must be re-emphasized that the noseless model $\bA_{\kappa+1} $ should be used to compute the PCRLB using Algorithm \eqref{alg:pcrlb} --- otherwise the computed PCRLB won't be as accurate.  
From this, it is also true that the PCRLB cannot be computed in real-time applications where the true model $\bA_{\kappa+1} $ is not known; however, the simulation analysis presented in Section \ref{subsec:RecMCdemo}, 
where PCRLB is compared to different filter covariance values, sheds new insights into system identification of equivalent circuit models. 

\begin{algorithm}[h]
\begin{footnotesize}
\caption{\\
\label{alg:pcrlb}
$[\bP_{\kappa+1} ]  ={\rm PCRLB}{ \left( \bP_\kappa, \bA_{\kappa+1} , \bSigma \right)}$
}
\label{alg:PCRLB}
\end{footnotesize}
\begin{algorithmic}[1]
\State{Innovation Cov: $ \bS_{\kappa+1}=\rm \bSigma +\bA_{\kappa+1}\bP_\kappa\bA_{\kappa+1}^T $}
\State{Filter Gain: $ \bW_{\kappa+1}=  \bP_\kappa\bA_{\kappa+1}^T  \bS_{\kappa+1}^{-1}$}
\State{State Est. Cov:  
$\bP_{\kappa+1}=  \bP_\kappa- \bW_{\kappa+1}\bS_{\kappa+1}\bW_{\kappa+1}^T  $ }
\end{algorithmic}
\end{algorithm}

\begin{remark}
\end{remark}
For the sake of generality, in this section, we treated as though the parameter $\bb_\kappa$ is a vector. 
In the next section, during the numerical evaluations of the proposed algorithms in this section, we will go back to considering the case where $\bb = R$, i.e., a scalar. 
The generic derivations presented in this section will be useful to understand more realistic models, such as the ones presented in \cite{BFGpart12014} for battery equivalent circuit model parameter estimation. 

\begin{remark}
\end{remark}
{It} can be verified (using matrix inversion lemma) that the the updated covariance matrix computed using Algorithm \ref{alg:PCRLB} and \eqref{eq:InfoUpdate} are the same.  
The goal of displaying Algorithm \ref{alg:PCRLB} is to stress the importance of PCRLB and to differentiate it from CRLB shown in \eqref{eq:crlb-kappa}. 

\begin{remark}
\end{remark}
When there is no prior information, the PCRLB will be identical to the CRLB. 
In this paper, no such prior information is assumed; as such, the terms PCRLB and CRLB are used interchangeably in the remaining sections. 

\subsection{Monte-Carlo Demonstration}
\label{subsec:RecMCdemo}
The goal of this subsection is to compare the performances of the recursive filtering approaches summarized in Section \ref{sec:RecusiveEST}.
Once again, we will adopt a Monte-Carlo simulation approach to compare performance where the following features of the simulated data remains the same to that in Section \ref{sec:MCdemo_LSvsTLS} for one batch of observation in \eqref{eq:vk}:
\begin{itemize}
\item The true current is kept constant at $i(k) = i_c = 2 \,\, {\rm A}$ and the true resistance is also kept constant at $R = 0.25 \,\,  \Omega$
\item
The definition of SNR is the same as in \eqref{eq:SNR}
\end{itemize}
Using the above definition and the observation model defined in \eqref{eq:vk},
the voltage measurement $\bz_{\rm v}(\kappa) = \bz_{\rm v}$
and
current measurement $\bz_{\rm i} = \bA_\kappa$ are generated for different SNR values ranging from 0 dB to 60 dB according to the model defined in \eqref{eq:noisymodel}--\eqref{eq:ni(k)} under the assumptions in \eqref{eq:assumptions}, for 
$\kappa = 1, 2, \ldots, N$ consecutive batches. 
The simulated measurements $\bz_{\rm v}(\kappa), \,\, \kappa = 1,2, \ldots, N,$ were used to estimate the parameter $\bb = R$ using the following algorithms: 
\begin{itemize}
\item[(a)]
The RLS estimator (summarized in Algorithm \ref{alg:RLS}). 
\item[(b)]
The TLS estimator (summarized in Algorithm \ref{alg:RTLS}). 
\item[(c)]
The TKF estimator (summarized in Algorithm \ref{alg:RTLS-KF}). 
\end{itemize}
The simulation was repeated for 1000 Monte-Carlo runs in order to computed the following performance indicators for each of the above estimators:
\begin{itemize}
\item 
The average estimate (instead of the bias in previous examples)
\item 
The SDE - see it defined for the LS estimator in \eqref{eq:NORMvar}
\end{itemize}
Using the knowledge of the noiseless model $\bA_\kappa$,  the PCRLB is computed in order to compare the performance of the 
above estimators with the theoretical lower bound. Observations from these results are presented in the remainder of this subsection.

\subsubsection*{Comparison of RLS and TLS}
\begin{figure*}
\begin{center}
\subfloat[][]
{\includegraphics[width=.95\columnwidth]{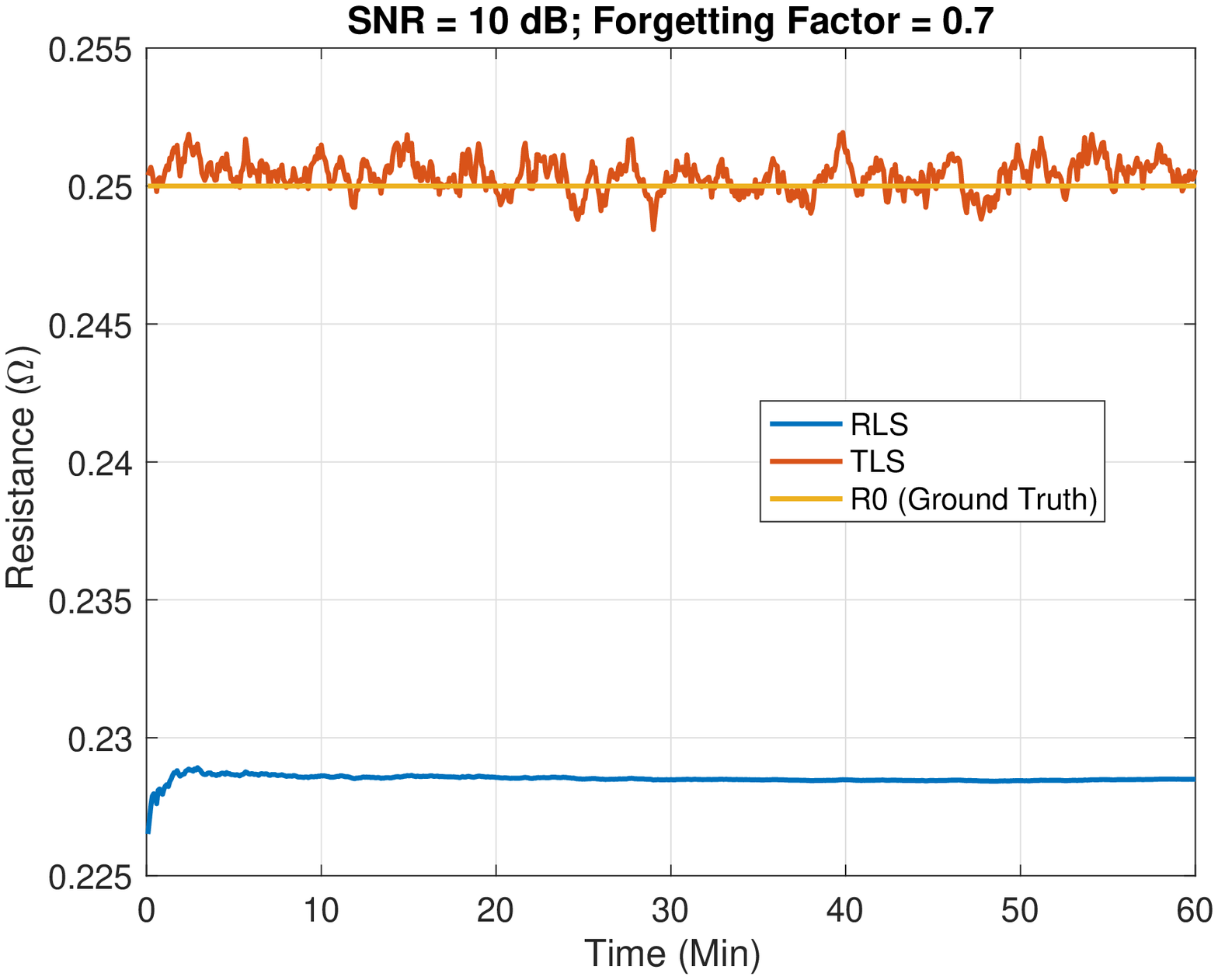}} \hspace{12 pt}
\subfloat[][]
{\includegraphics[width=.95\columnwidth]{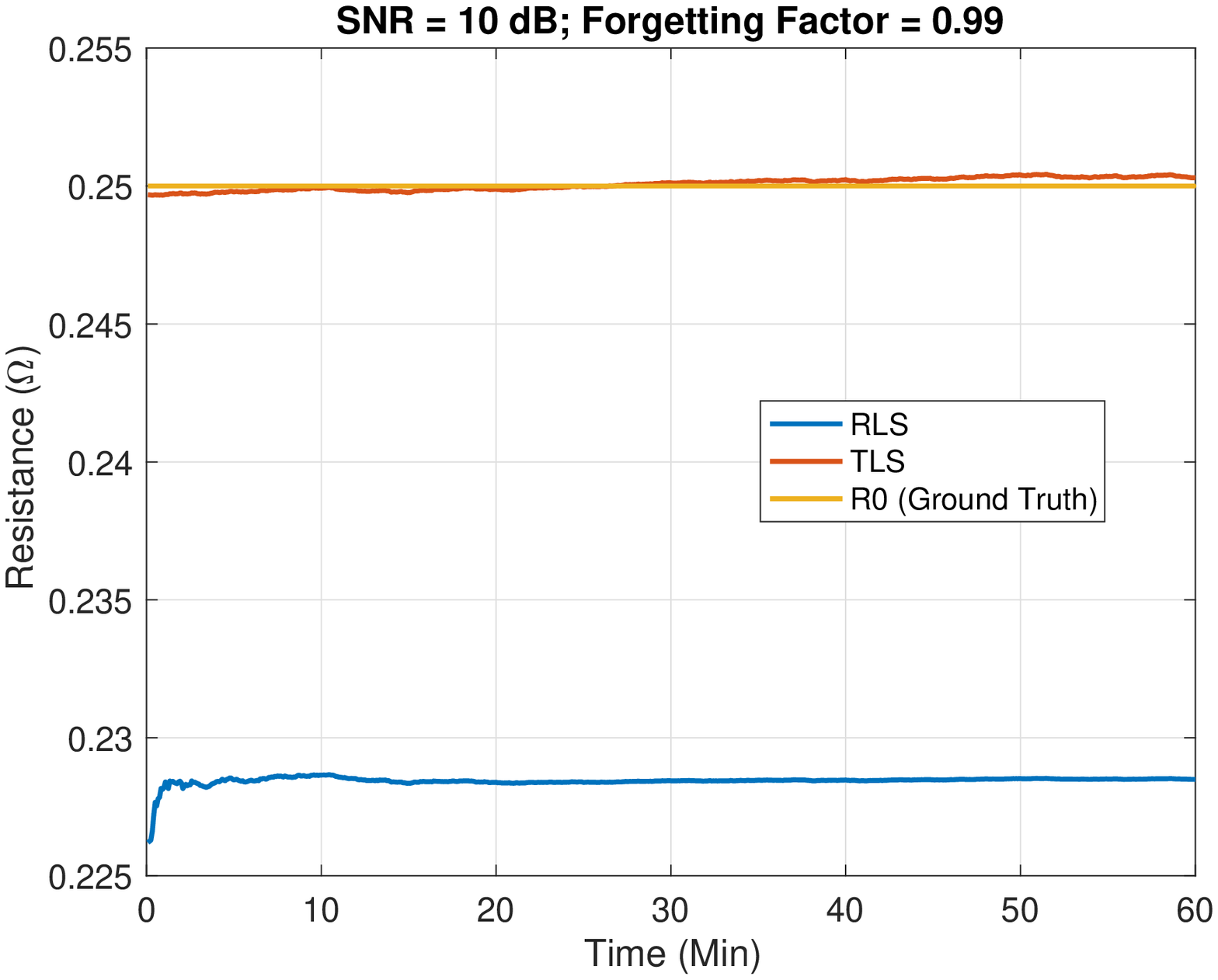}} \\
\subfloat[][]
{\includegraphics[width=.95\columnwidth]{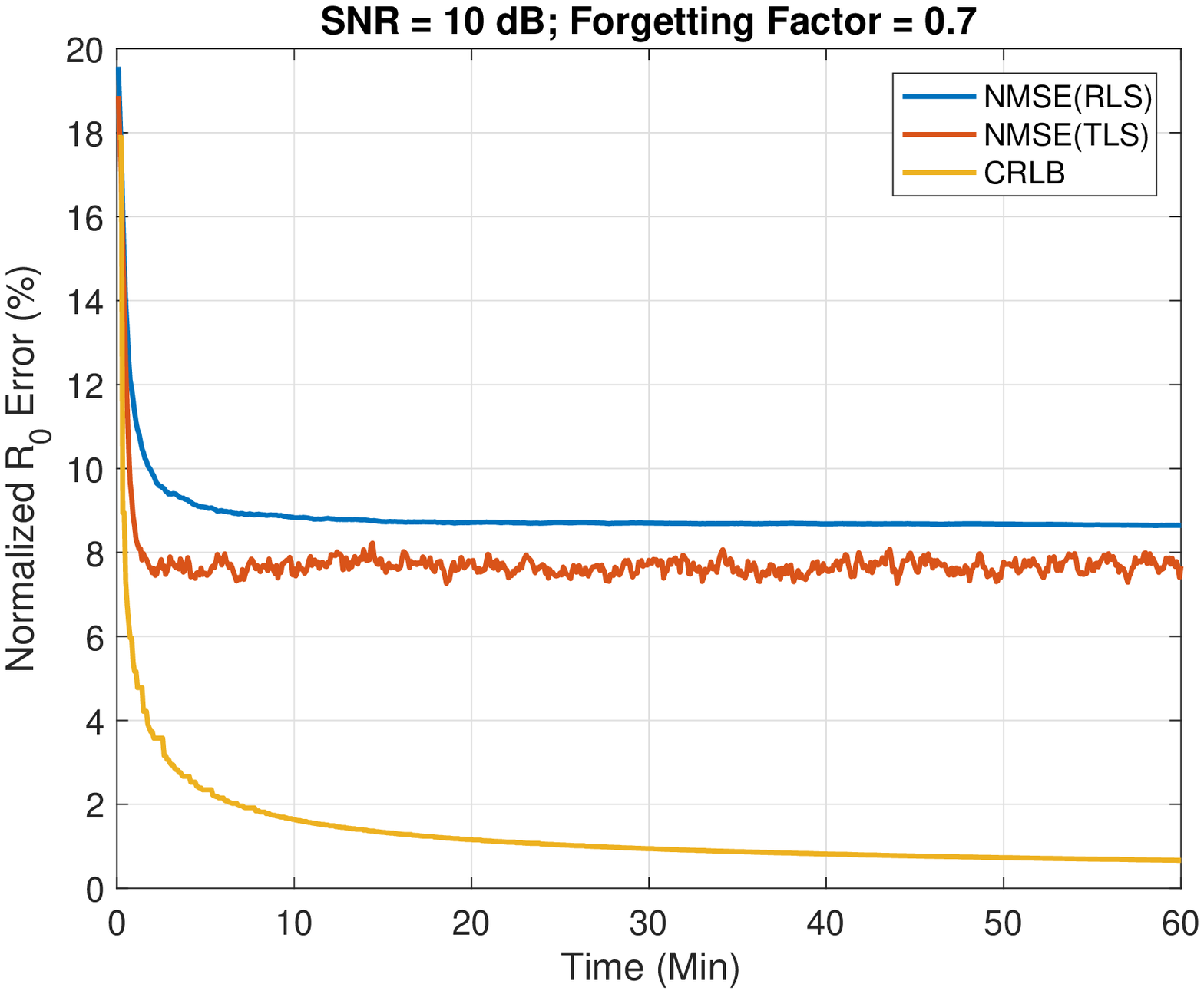}} \hspace{12 pt}
\subfloat[][]
{\includegraphics[width=.95\columnwidth]{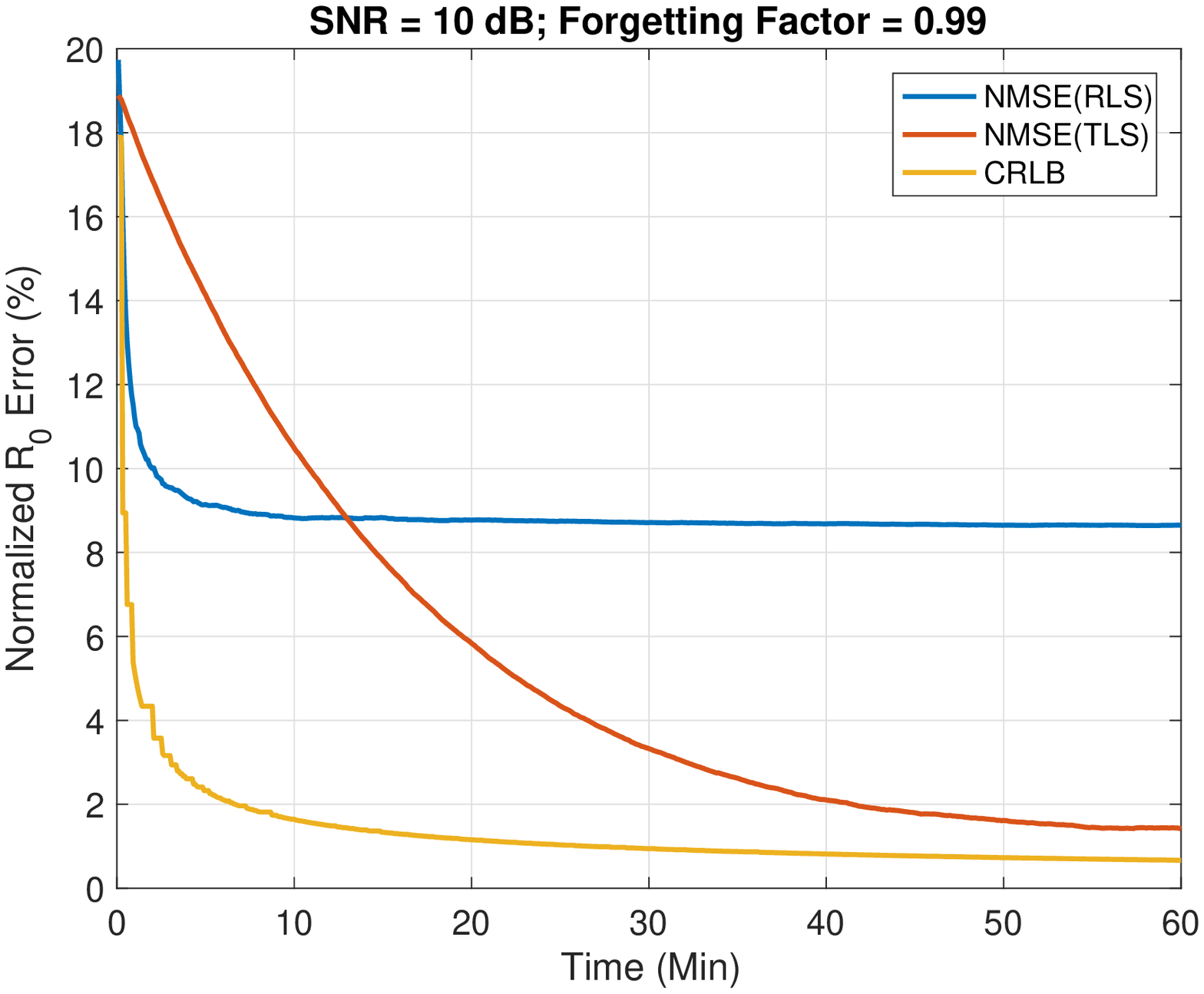}}
\end{center}
\caption{
{\bf Performance comparison of the RLS and TLS estimators. }
\textbf{\em Left column:} when the forgetting factor is $\lambda = 0.7$.
\textbf{\em Right column:} when the forgetting factor is $\lambda = 0.99$.
\textbf{\em Top row:} true vs. averaged estimate over 1000 Monte Carlo runs. 
\textbf{\em Bottom row:} normalized standard deviation of error in comparison to (normalized) CRLB.  
This figure demonstrates the trade-off of the recursive total least square (TLS) filter with respect to the forgetting factor $\lambda:$
when the forgetting factor is low (left column) the TLS filter converges fast for a high variance of estimation;
when the forgetting high (right column) the TLS filter converges slow, but the variance of estimation reaches the CRLB. 
All plots in this figure, except the CRLB, are averaged over 1000 Monte Carlo runs. 
}
\label{fig:RecursiveDemo1}
\end{figure*}

Figure \ref{fig:RecursiveDemo1} summarizes the performance comparison of the RLS filter and the TLS filter. 
The top (row) shows the true parameter $R$ against the estimated ones and the bottom row shows the normalized estimation error along with the PCRLB (which is also normalized). 
The top row shows that the RLS estimator has a significant bias: the true parameter is $R = .25 \Omega $ whereas  the RLS estimator converges to a value of $\hat R_{\rm RLS} = .228\Omega $; the top row clearly shows the bias problem in the RLS estimator in the presence of model uncertainty. 
The top row also shows that the TLS estimator is better and is nearly unbiased compared to the RLS estimator. 
The bottom row in Figure \ref{fig:RecursiveDemo1} compares the normalized standard deviation of error of the RLS estimator (defined above) with that of the TLS filter and the normalized CRLB. 

In order to demonstrate the effect of the forgetting factor on the performance the left column had a forgetting factor of $\lambda = 0.7$ and the right column had $\lambda = 0.99$.
The bottom row shows the tradeoff of the TLS estimator: with smaller forgetting factor $\lambda$, the TLS estimator {\em converges fast} but at the cost of higher {\em variance of the estimated parameter};
with higher forgetting factor $\lambda$, the TLS estimator takes much longer to converge, but accrues much less variance of the error in the estimate. 
It appears that with a high enough value for $\lambda$ and longer time, the TLS estimates might become efficient, i.e., the variance of the estimator will be the same as the CRLB. 
This trade-off motivates one to seek an approach that can simultaneously retain the nearly unbiased nature and low variance of the TLS estimator. 

\subsubsection*{Demonstration of total Kalman filter}
\begin{figure}
\begin{center}
\subfloat[][]
{\includegraphics[width=.9\columnwidth]{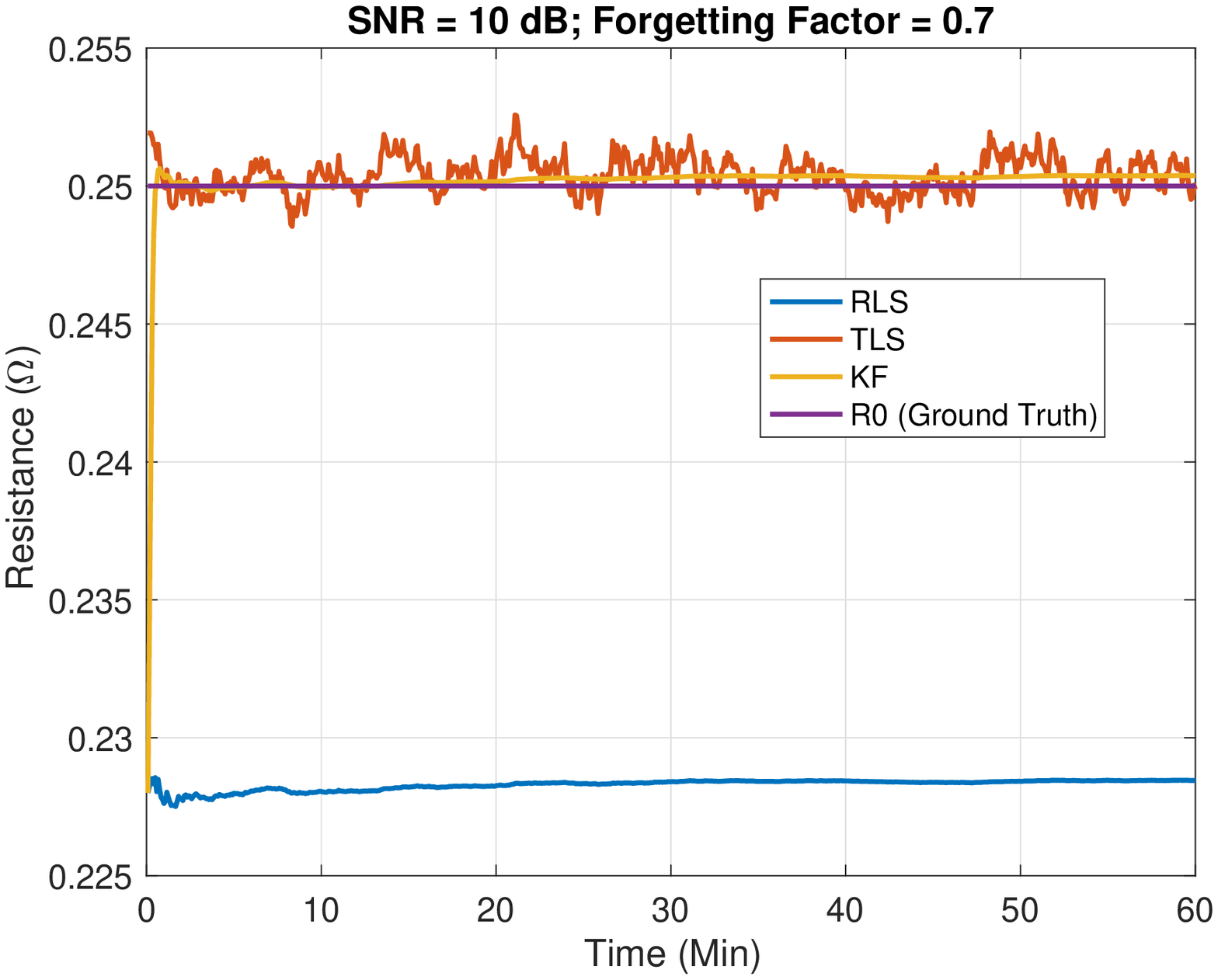}}\\
\subfloat[][]
{\includegraphics[width=.9\columnwidth]{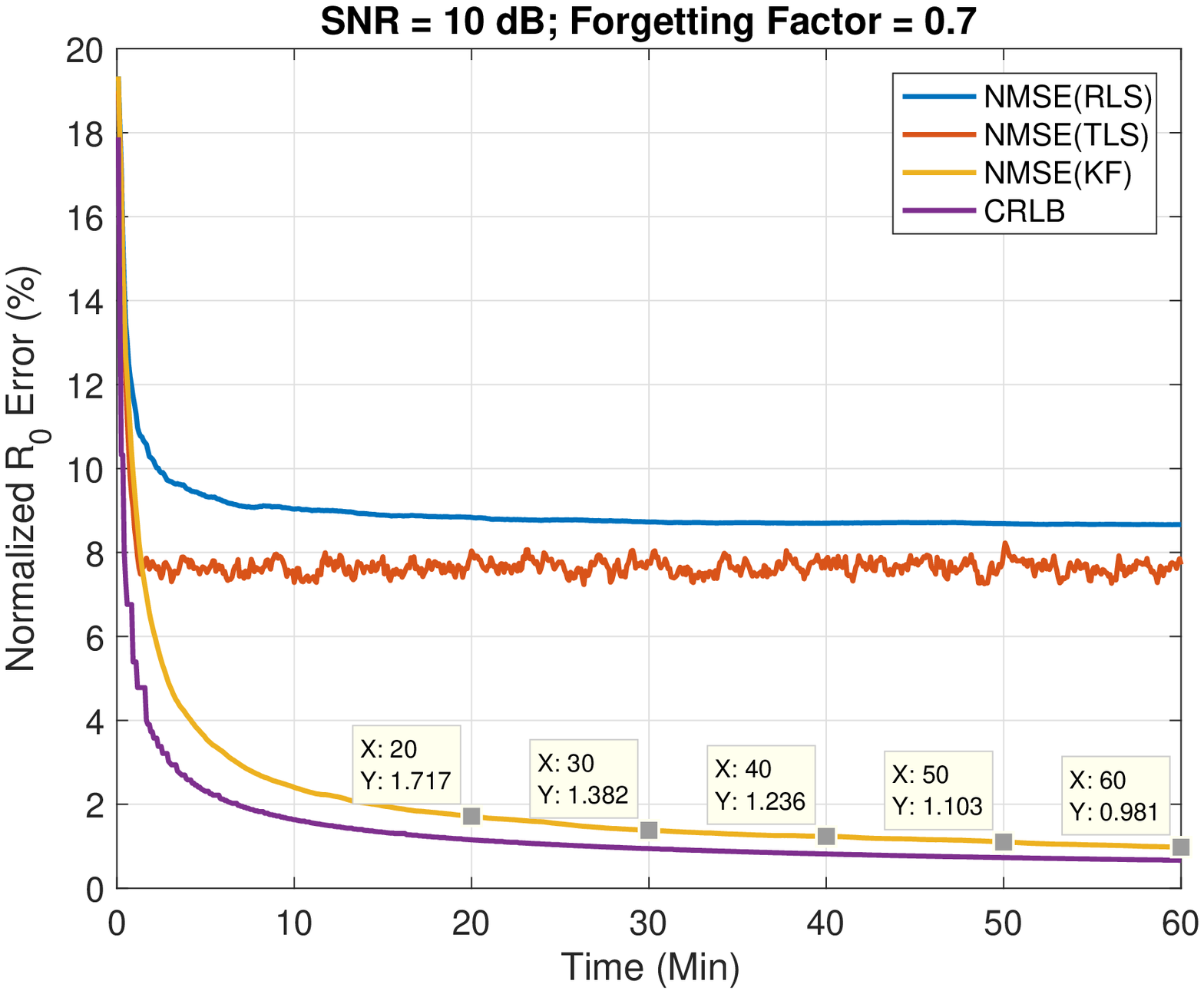}}
\end{center}
\caption{
{\bf Performance comparison of the total Kalman filter (TKF). }
Top plot shows the true parameter along with the estimated ones by RLS, TLS and TKF estimators.
The plot at the bottom compares the filter variance with computed CRLB. 
}
\label{fig:TLS-KF-demo}
\end{figure}

Figure \ref{fig:TLS-KF-demo} shows the performance of the TKF estimator with that of RLS and TLS estimators.
The top plot compares the true parameters against averaged values of the estimates over 1000 Monte Carlo runs. 
The observation here is that due to the bias of the RLS estimator, the TLS and TKF filters significantly outperform it. 
Further, the TKF estimates show less variance compared to the TLS estimates.
Figure \ref{fig:TLS-KF-demo}(b) shows the comparison of variance of all three estimators with CRLB;
it shows that the TKF is able to converge faster to a much lower variance of error in the estimated parameter compared to the TLS filter. 

\section{Time-varying SNR}
\label{sec:time-var-snr}

\begin{figure}
\begin{center}
\includegraphics[width=.9\columnwidth]{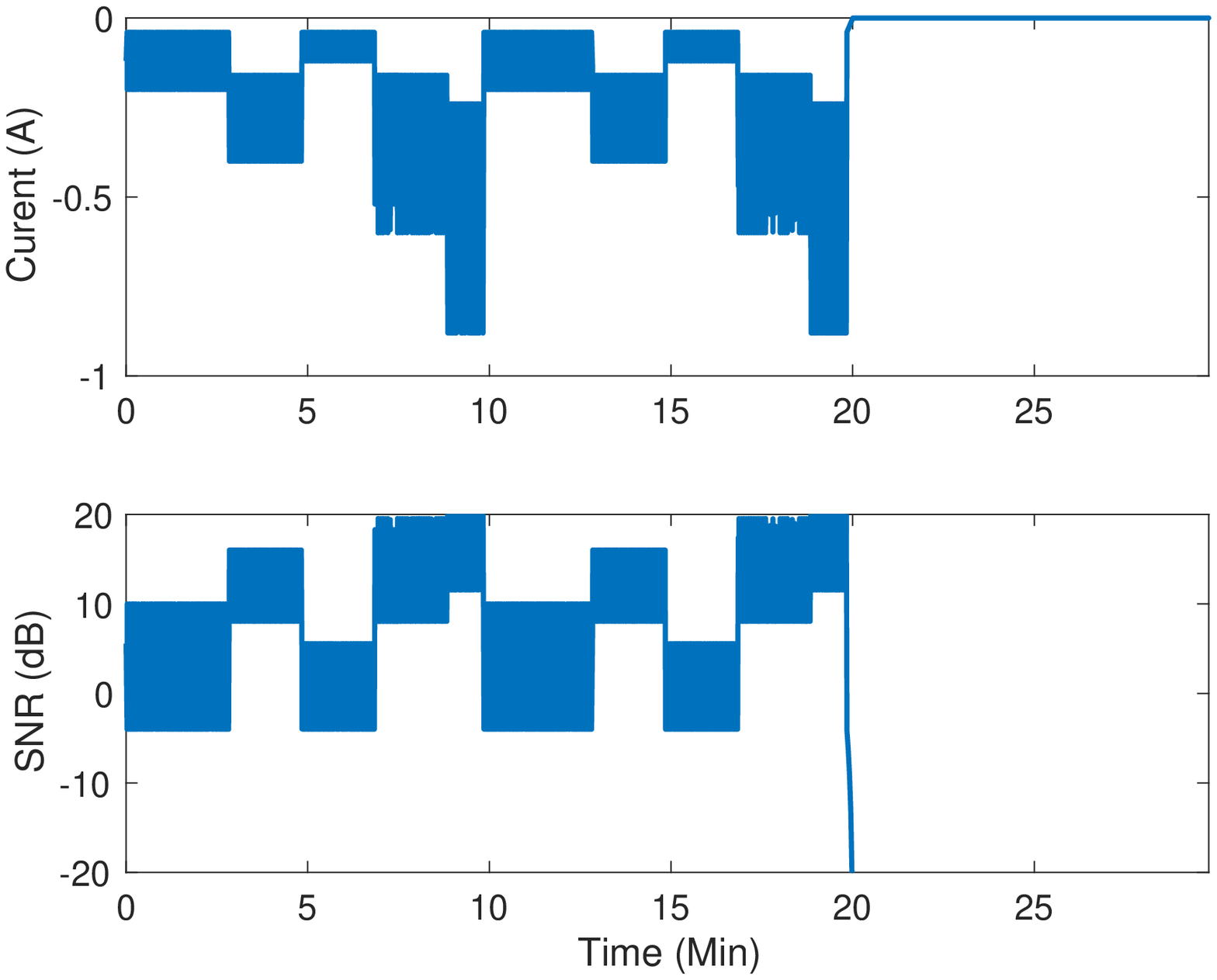}
\end{center}
\caption{
{\bf Time varying signal-to-noise ratio. }
The top plot shows typical current through a smart phone battery and the plot at the bottom shows the corresponding SNR assuming a current measurement noise standard deviation of $\sigma_i = 20 \, {\rm mA.}$
}
\label{fig:dynSNR}
\end{figure}

So far, we considered examples where the true current is a constant, i.e. $i(k) = i_c$. 
However, in most real-time applications, the current will not be a constant. 
For example, let us consider the application of battery management systems in electrical vehicles, where it is important to know the remaining power at any given time.
This requires accurate estimation of battery's internal impedance based on the current through the battery, while the vehicle is on the road where the current through an electric vehicle ultimately depends on the driver.
In this section, we will analyze the performance of the algorithms from Section \ref{sec:RecusiveEST} when the current $i[k]$ is time varying. 

The top plot in Figure \ref{fig:dynSNR} shows typical current that was simulated based on realistic parameters \cite{avvari2015experimental} through a smart phone. 
It can be noticed that with time varying current, the SNR is also a time varying:
\begin{align}
{\rm SNR}(k) = 20 \log_{\rm 10} \left( \frac{i(k) R}{ \sqrt{\sigma_{\rm v} ^2 + \sigma_i^2 R^2 }} \right)
\end{align}
The plot at the bottom in Figure \ref{fig:dynSNR} shows the corresponding ${\rm SNR}(k)$.
It can be observed that the SNR fluctuates between positive and negative values. 
When the current is zero, starting from 20 min. until 30 min., the SNR drops to negative infinity.

First let us consider the progression of CRLB with time-varying SNR. 
Under the scalar parameter assumptions in \eqref{eq:assumptions}. 
The CRLB at the end of the $\kappa^{\rm th}$ batch of lengh $m$ is given as 
\begin{align}
 \sigma_{\rm CRLB}^2(\kappa) & = \sigma_{\rm v} ^2 \left({\sum_{k=1}^{m\kappa }} i(k)^2 \right)^{-1}
 \label{eq:crlb}
\end{align}
It must be noted that such a closed form expression is not possible for the generic model \eqref{eq:vk}.
However, it can be recursively computed using Algorithm \ref{alg:pcrlb}.

\begin{figure*}
\begin{center}
\subfloat[][30 minute current profile]
{\includegraphics[width=.95\columnwidth]{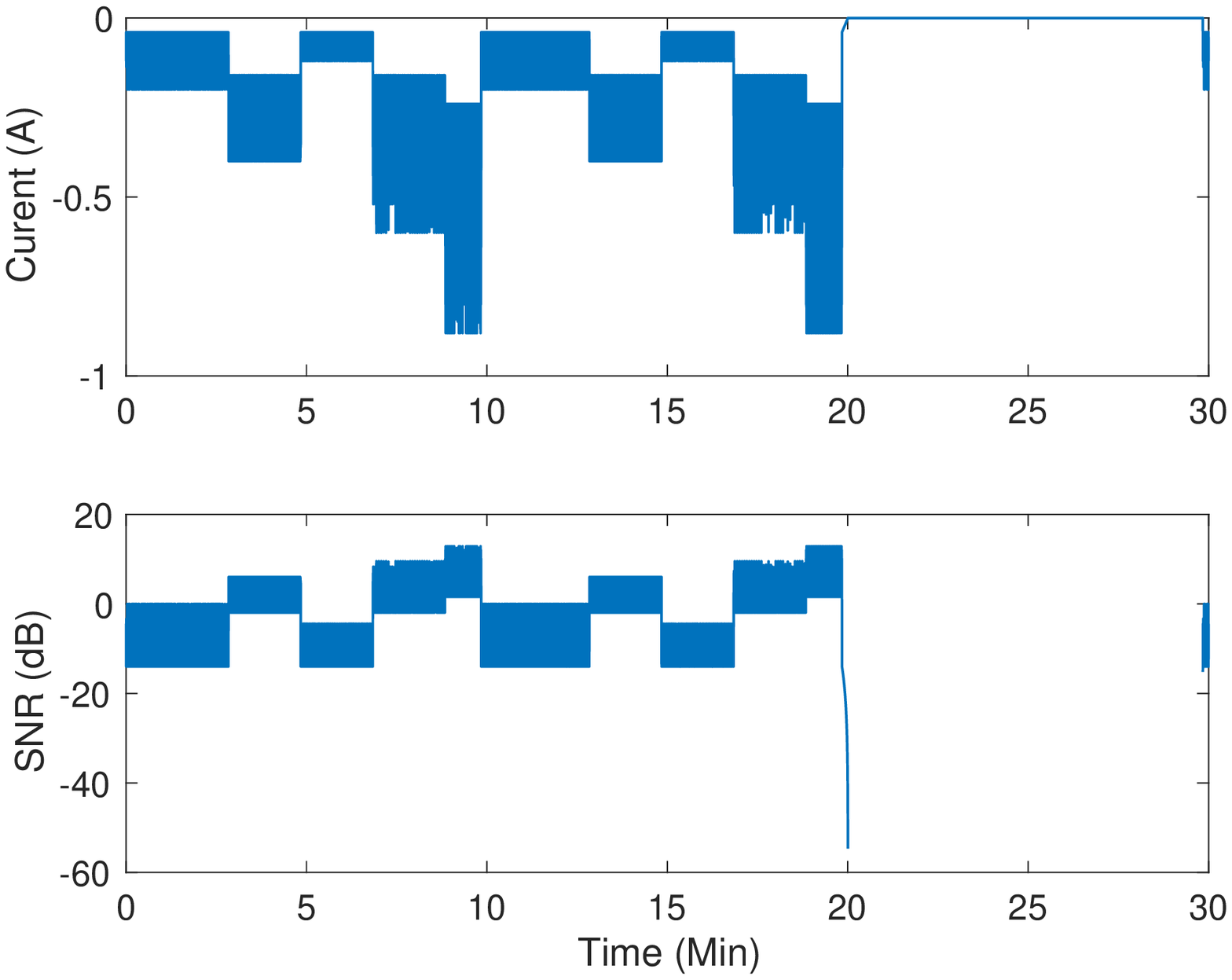}} \hspace{12 pt}
\subfloat[][12 cycles of 30 minute current profile shown on the left]
{\includegraphics[width=.95\columnwidth]{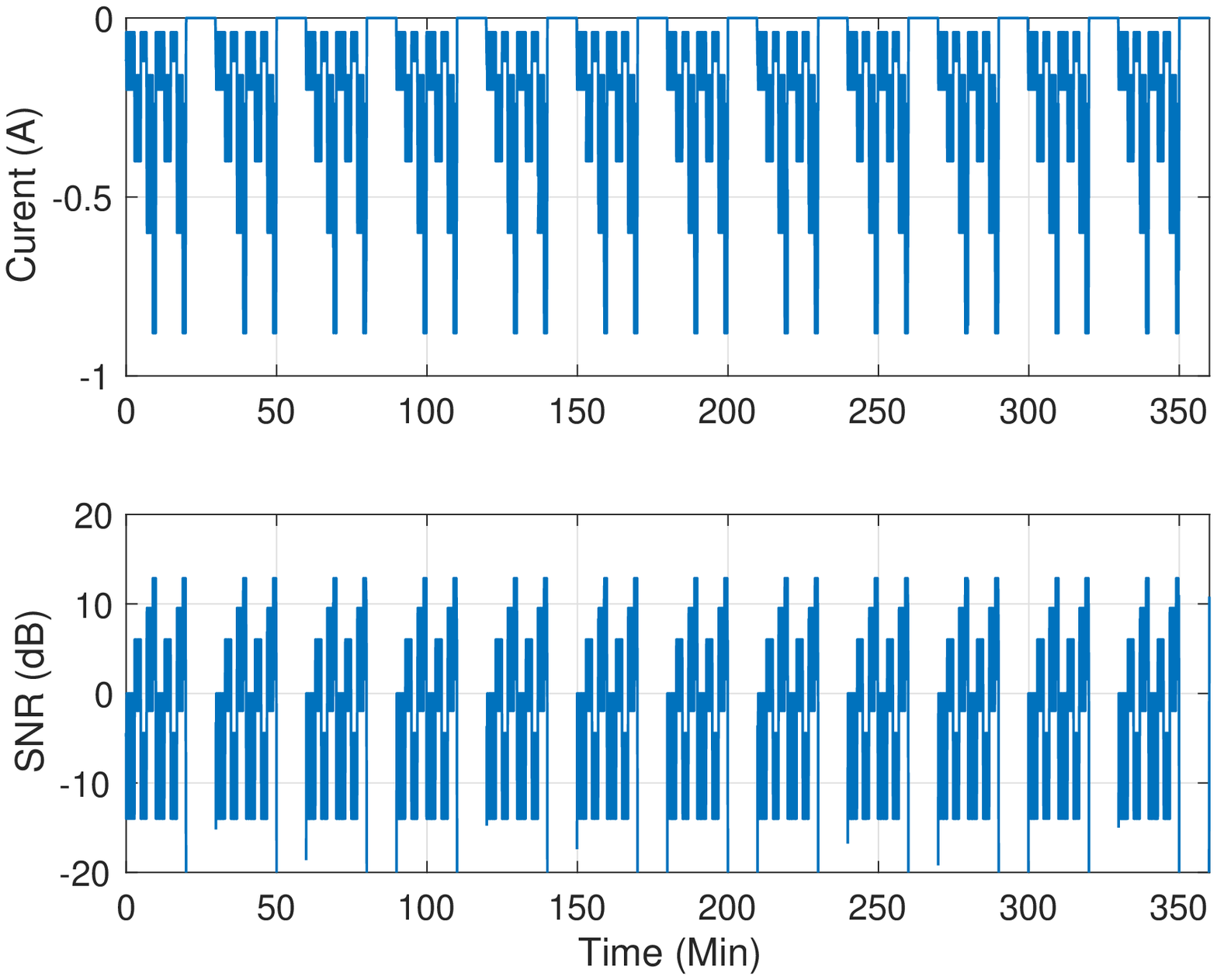}}\\
\subfloat[][PCRLB (plot at the bottom is zoomed version)]
{\includegraphics[width=.95\columnwidth]{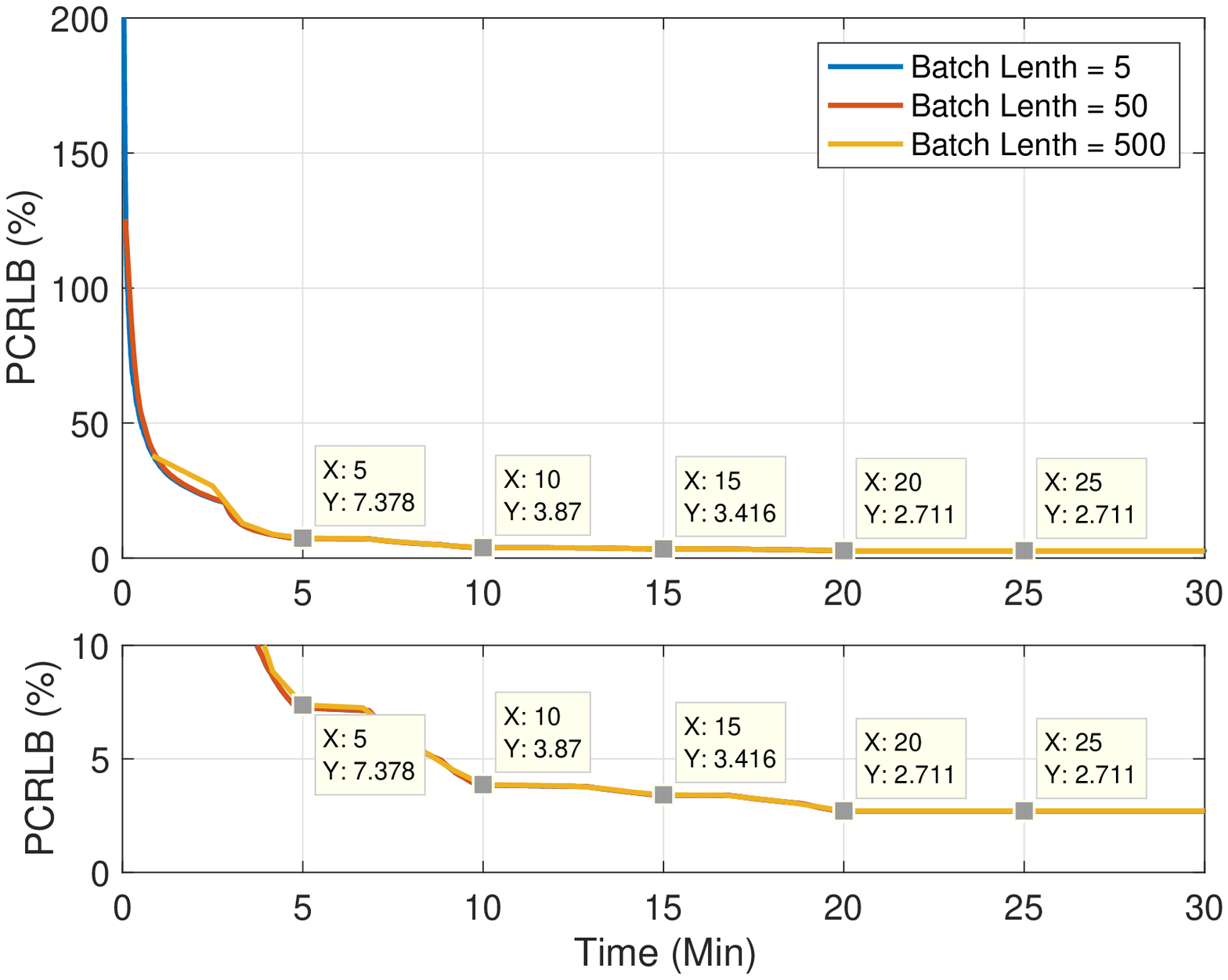}} \hspace{12 pt}
\subfloat[][PCRLB (plot at the bottom is zoomed version)]
{\includegraphics[width=.95\columnwidth]{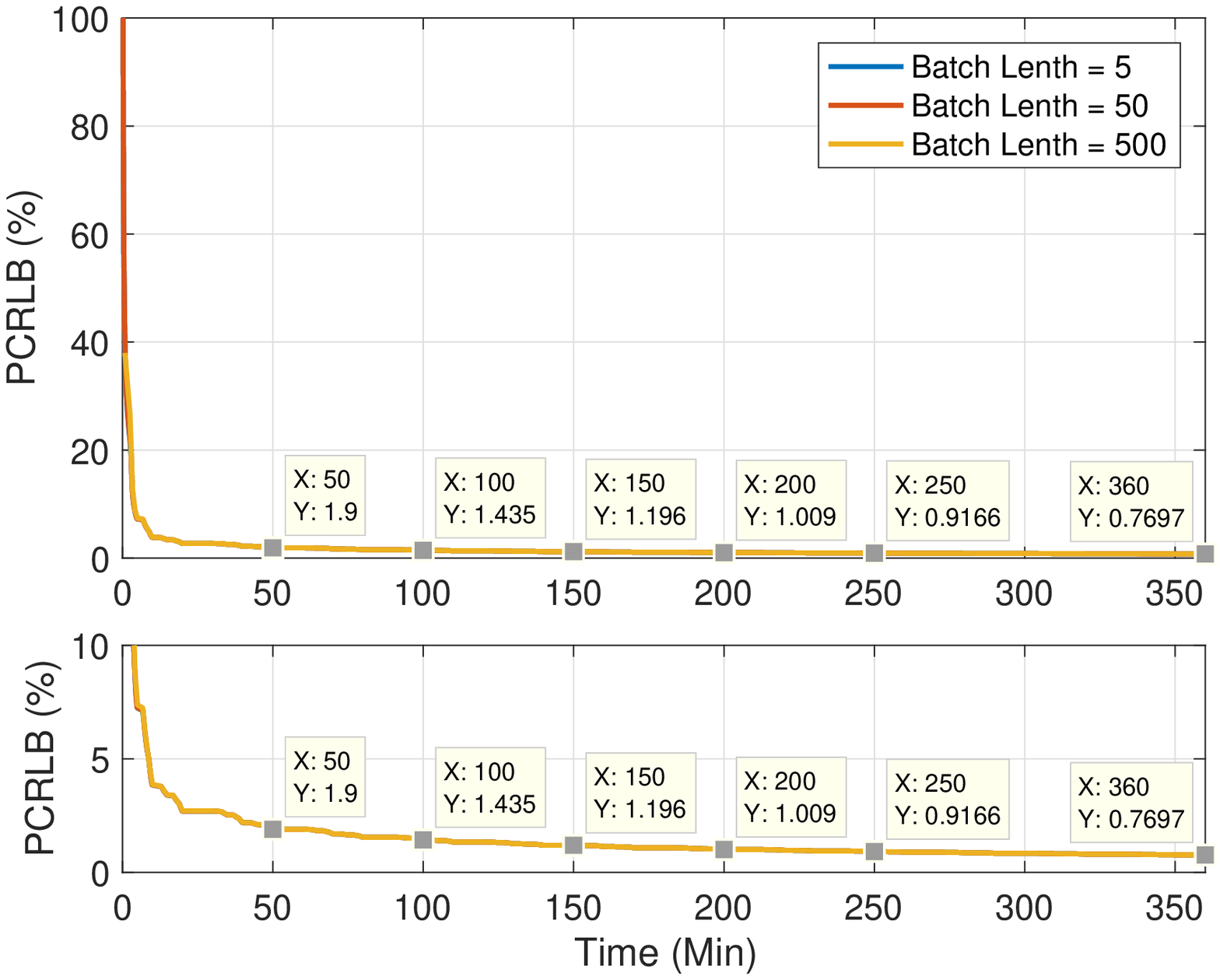}}
\caption{
{\bf Normalized PCRLB during time-varying SNR.}
The plots show that the theoretical error bound decrease as more information is received. 
When there is no new information, i.e., during the period of zero current, the PCRLB remains the same; this gives a guideline to the expected behaviour of ECM parameter estimators. 
\textbf{\em Left-column.} 
30-minute load profile containing zero-current for the last 10 minutes.
\textbf{\em Right-column.} 
360-minute load profile that is made up by replicating the profile on the left for 12 times. 
\textbf{\em Top-row.} 
Current profile and the corresponding SNR. 
\textbf{\em Bottom-row.} 
The PCRLB.
}
\label{fig:PCRLB(k)}
\end{center}
\end{figure*}

Figure \ref{fig:PCRLB(k)} shows the PCRLB against time for dynamic current profile shown in Figure \ref{fig:dynSNR}. 
The left column of the figure shows a 30-minute dynamic current profile (also shown in Figure \ref{fig:dynSNR}) and the corresponding CRLB -- this is intended to provide a closer look at how PCRLB changes with time varying signals. 
On the right column, a time-varying signal of longer duration and the corresponding PCRLB is displayed. 
It is important to note that the PCRLB is a performance measure of consecutive data, i.e., it takes into account all the preceding information and provides the performance bound of a filter at the present time instance. 
For example, consider the zero-current period, from 20 minute mark to 30 minute mark in Figure \ref{fig:PCRLB(k)}(a).
Since there is no information, one would expect infinite error; however, considering the prior information, from 0 minute mark to 20 minute mark, the performance bound is finite - PCRLB provides the exact value of this bound. 

\begin{figure}
\begin{center}
\subfloat[][Averaged over 1000 Monte Carlo runs]
{\includegraphics[width=.85\columnwidth]{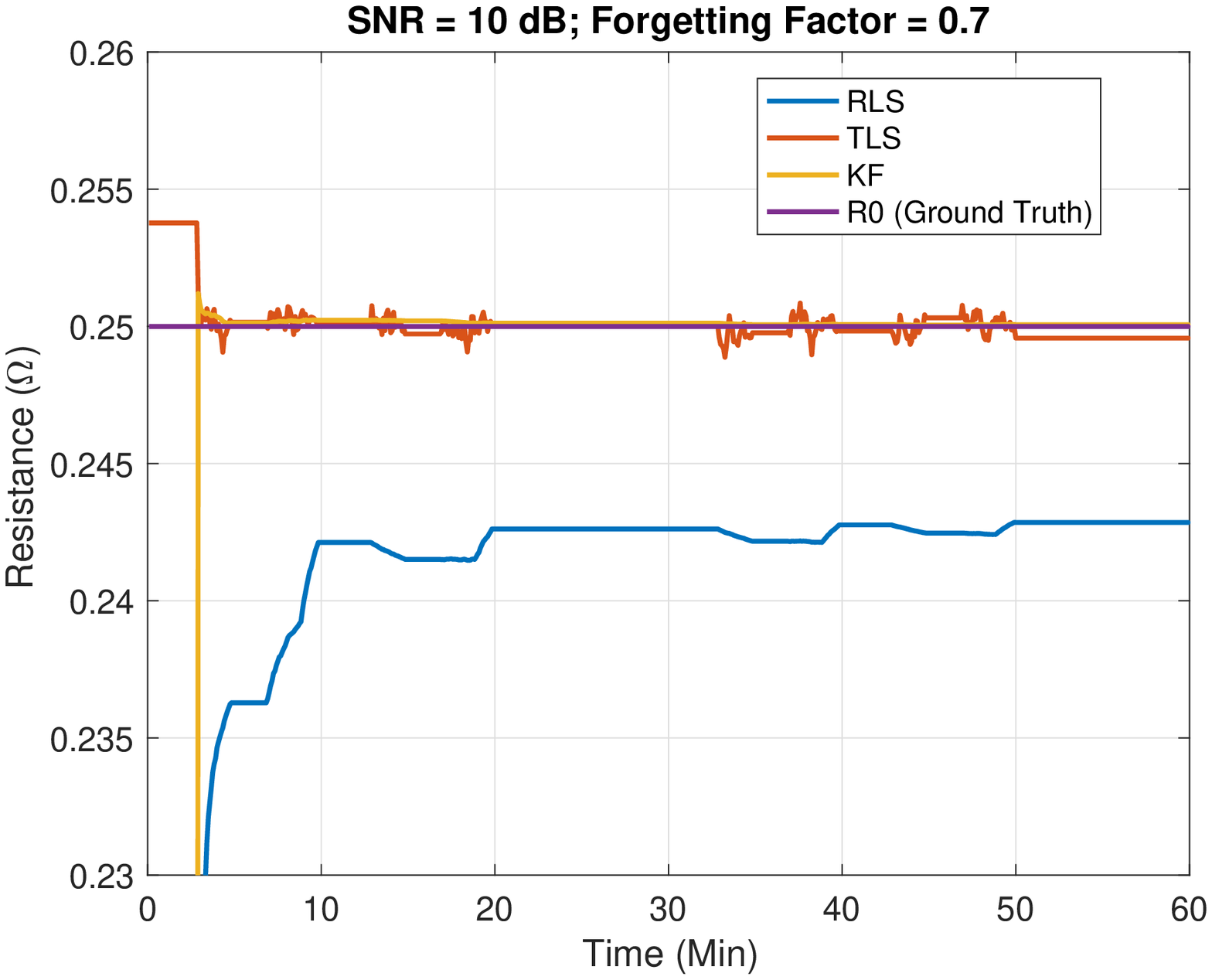}}  \hspace{12 pt}\\
\subfloat[][Averaged over 1000 Monte Carlo runs]
{\includegraphics[width=.85\columnwidth]{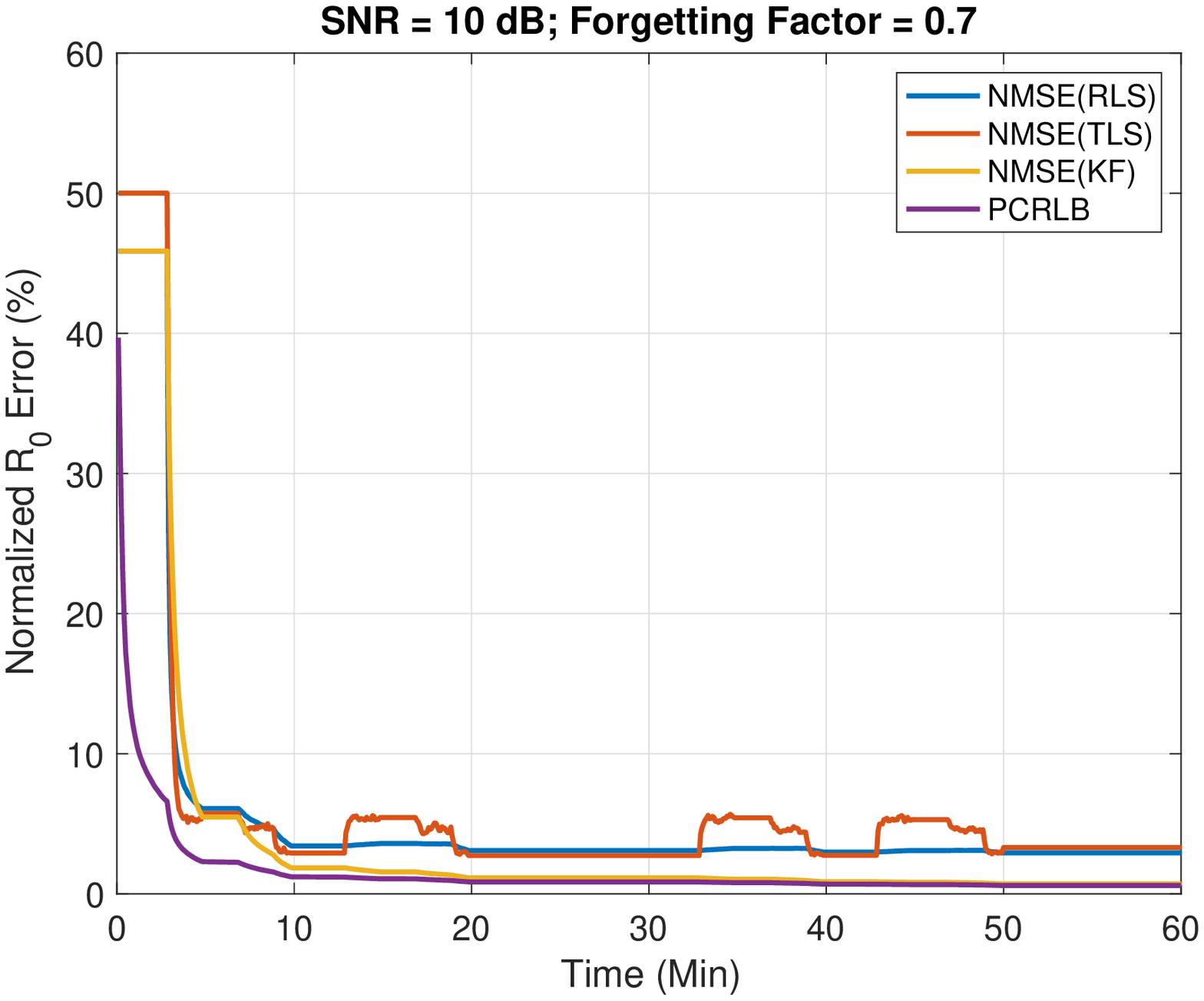}} \hspace{12 pt}
\end{center}
\caption{
{\bf Performance comparison during time-varying SNR. }
{\em Top row:} True vs. estimated parameter by different estimators.
{\em {Bottom row:}} Normalized standard deviation of error by different estimators.}
\label{fig:Recursive1run}
\end{figure}

Having analyzed the performance bound using the CRLB, let us now analyze the performance of the three estimators, RLS, TLS, and TKF, for a dynamic SNR scenario. 
The followoing parameters were used throughout the analysis:
\begin{itemize}
\item
A 60 minute portion of the dynamic load profile shown in Figure \ref{fig:PCRLB(k)} is used for the time varying current $i(k)$.
\item The true resistance is kept constant at $R = 0.25 \,\,  \Omega$
\item
The current and voltage measurement noise are generated using the following parameters:
$\sigma_i = 200 {\rm mA}$ and $\sigma_{\rm v}  = 200 {\rm mV}$ 
\end{itemize}
Using the above definition and the observation model defined in \eqref{eq:vk} under the assumptions \eqref{eq:assumptions},
the voltage measurement $\bz_{\rm v}(\kappa) = \bz_{\rm v}$
and
current measurement $\bz_{\rm i} = \bA_\kappa$ are generated according to the model \eqref{eq:vk} for $\kappa = 1, 2, \ldots, N=720$ consecutive batches (for a sampling time of 100 ms and total time duration of 1 hour). 
The simulated measurements $\bz_{\rm v}(\kappa), \,\, \kappa = 1,2, \ldots, 720,$ were used to estimate the parameter $\bb = R$ using the following algorithms: 
\begin{itemize}
\item[(a)]
The RLS estimator (summarized in Algorithm \ref{alg:RLS}). 
\item[(b)]
The TLS estimator (summarized in Algorithm \ref{alg:RTLS}). 
\item[(c)]
The TKF estimator (summarized in Algorithm \ref{alg:RTLS-KF}). 
\end{itemize}  
Figure \ref{fig:Recursive1run} summarizes the  performance of the above three recursive estimators. 
All three recursive filters were implemented using a batch length of $m=50$ samples; 
Further, the TLS estimates needs to be avoided when the current is zero. 
Considering that there is always current measurement noise, a current threshold is set on the TLS filter for each batch of observations. 
When the current is not above this threshold, the TLS estimator outputs the previous estimate. 
A good threshold can be obtained by observing the {\em information content} in each batch of observations, i.e., 
based on \eqref{eq:LScov}, the information content in that observation can be approximately written as 
\begin{align}
{\rm INFO}(\kappa) = \bA_\kappa^T \bSigma^{-1} \bA
\end{align}
where it must be stressed that the measured of information is only approximate due to the fact that $ \bA_\kappa$ is an observed quantity; as such, the ${\rm INFO}(k) $ serves only as a an approximate guide to setting the threshold. 
The TLS filter is switched off when the information in a batch of data is less than a specific threshold. 
Further, the simulation was performed under the assumptions in \eqref{eq:assumptions}, as such ${\rm INFO}(\kappa) $ is a scalar. 
For generic parameter estimation cases, the threshold must be developed by considering the nature of the {\em information matrix} ${\rm INFO}(\kappa) $; 
this scenario is out of the scope of this present paper and left for a future discussion.  
Finally, it was observed that the TLS filter does not need a threshold when $\lambda$ is high; that is because large amount of information is retained from preceding batches;
however, this would lead to slow convergence as we demonstrated in Figure \ref{fig:RecursiveDemo1}.

Unlike the Monte-Carlo experiments presented so far, the true parameter $R_0$ is not available with the real data being experimented in this section.

\section{Applications and Future Work}
\label{sec:applications}

The proposed approach in this paper has important applications in system identification. 
In this section, we give a brief overview of the potential applications of the algorithms and analysis presented in this paper. 

\begin{figure}[h]
\begin{center}
\subfloat[][ECM of a generic system]
{\includegraphics[width=.45\columnwidth]{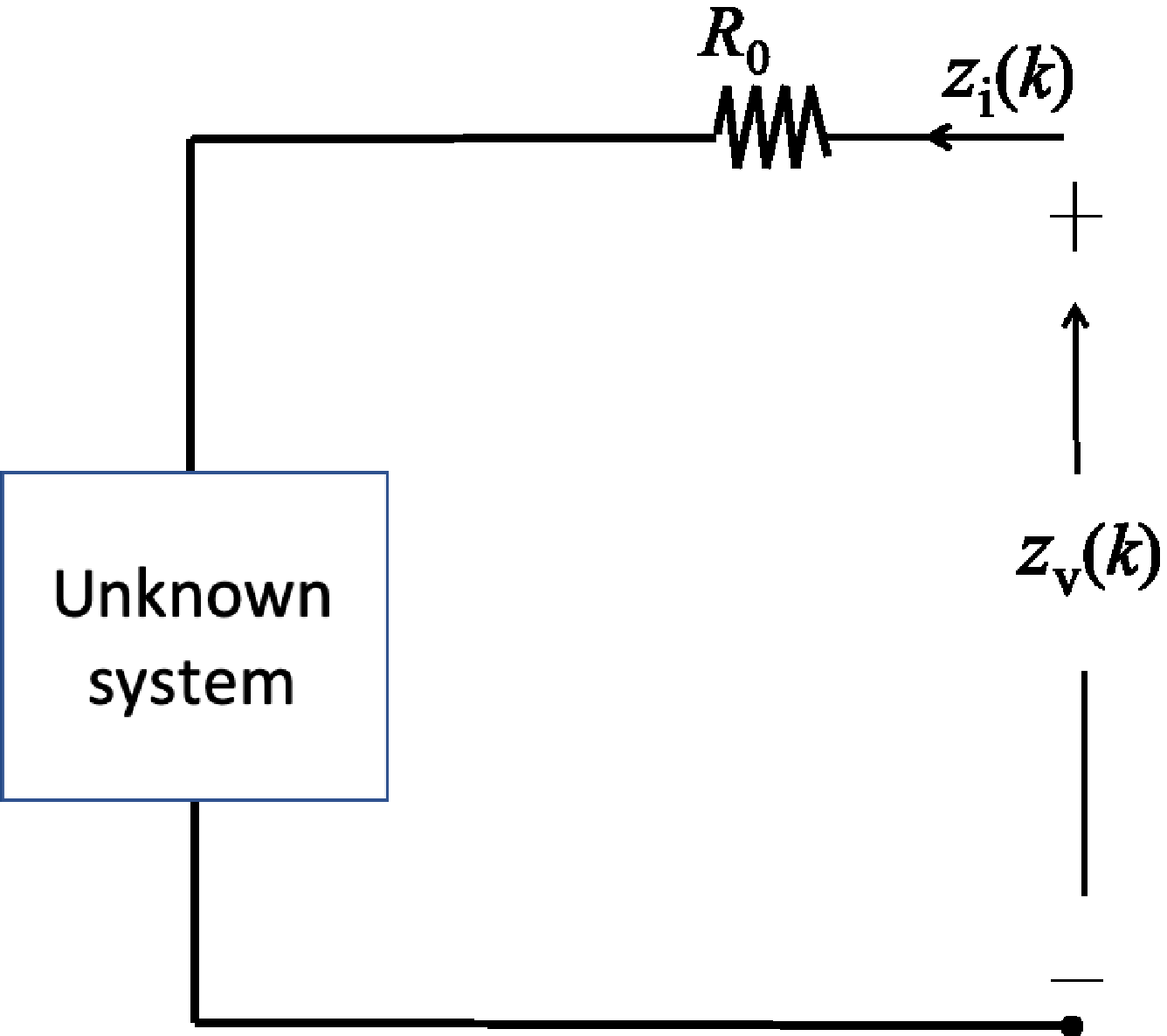}\label{(a)} }
\subfloat[][ECM of a battery]
{\includegraphics[width=.45\columnwidth]{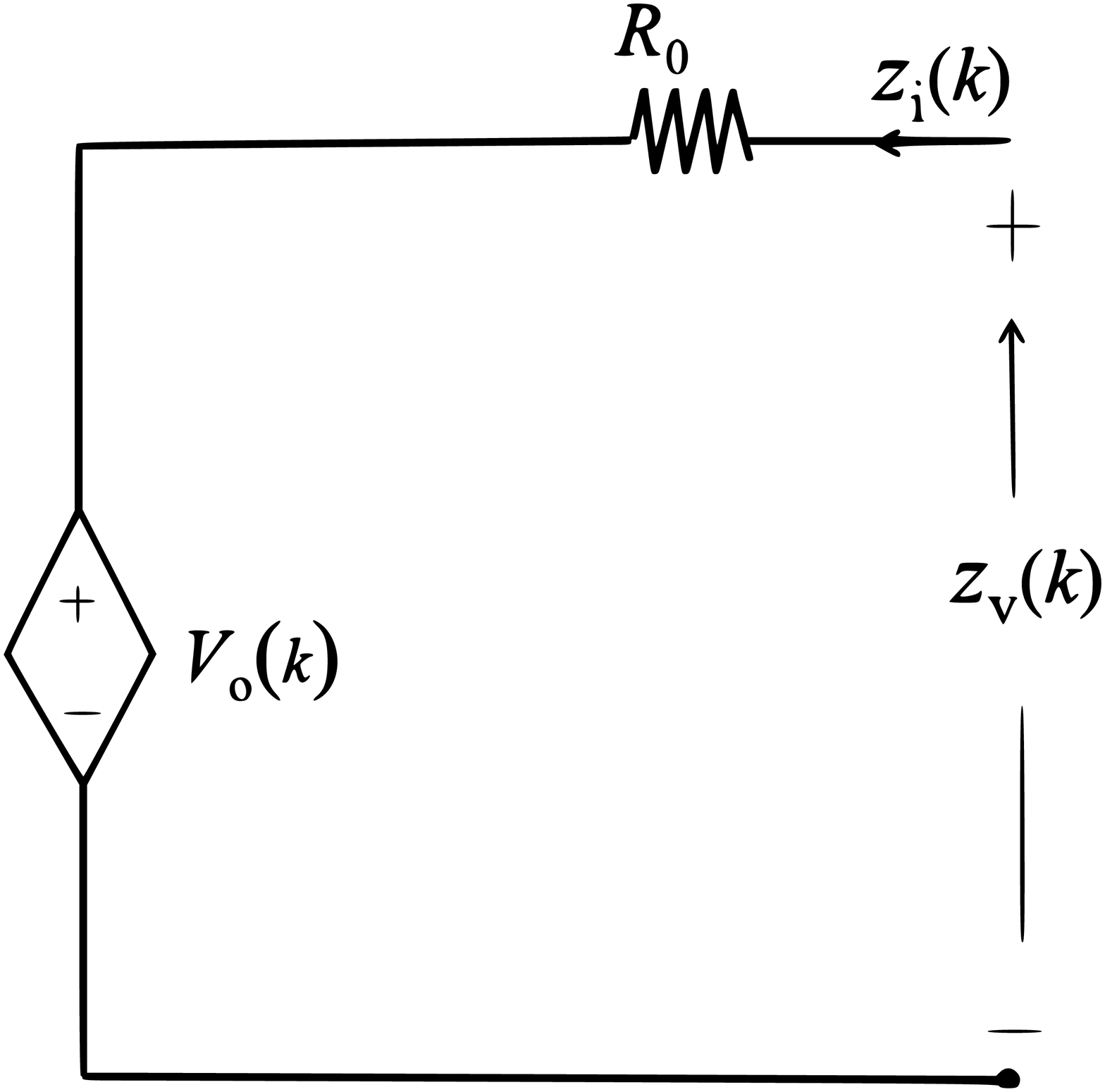}\label{(b)} }
\end{center}
\caption{
{\bf Application of the proposed methods in this paper.}
{Left figure:} shows an unknown system whose series resistance needs to be estimated based on the measured voltage and current as shown. 
{Right figure:} shown an example of such a system in rechargeable batteries. 
}
\label{fig:sysIDgeneric}
\end{figure}

It was assumed in this paper that the voltage across the ECM (in the present case ECM is just a resistor) and the current through it can be directly measured. 
Figure \ref{fig:sysIDgeneric}(a) shows a generic extension to this problem where an unknown system prevents the direct measurements of voltage across the resistor. 
A good example of such a scenario is found in Li-ion batteries \cite{aurilio2015battery,saha2008prognostics,schweighofer2003modeling,martinez2019development}, shown in Figure \ref{fig:sysIDgeneric}(b) where the battery electromotive force (EMF) represents the unknown system. 
Real-time battery impedance identification is an active research topic. 
The objective for the scenario in Figure \ref{fig:sysIDgeneric}(b) is to estimate the series resistance of the battery, $R_0.$ 
Observation models, similar to the one in \eqref{eq:vk} can be developed by leveraging the features of the unknown system \cite{BFGpart12014} as well as by ignoring it \cite{ArifEPEC2019}. 
Each approach has its own merits and drawbacks:
\begin{itemize}
\item[(a)]
{\em By considering the unknown system as a blackbox.}
Under this assumption, an input output model can be developed \cite{ArifEPEC2019} based on successive measurements of measured current, i.e., $z_{\rm i}(k), z_{\rm i}(k+1)$ and the voltage across the whole system, i.e., $z_{\rm v}(k), z_{\rm v}(k+1) $ by assuming that the voltage across the unknown system remains constant between two adjacent observations.
The attractive feature of this approach is that it does not require any knowledge about the unknown system. 
However, this approach observes the system in terms of current-difference $z_{\rm i}(k)-z_{\rm i}(k+1)$  and voltage difference $z_{\rm v}(k)-z_{\rm v}(k+1) $ --- this results in significant reduction to the SNR of the observation model. 
Prior sections in this paper gave detailed isights about how low SNR will result in poor performance.
\item[(b)]
{\em By assuming the knowledge of the unknown system.}
The EMF of the battery is represented as a voltage $V_\circ(k)$ that is related to the state of charge $s(k)$ of the battery.
For example, the combined model assumes that 
 \begin{align}
 V_\circ(k) = k_0 + \frac{k_1}{s(k)} + k_2s(k) + k_3 \ln(s(k)) + k_4\ln(1-s(k)) 
 \label{eq:combined}
 \end{align}
 Using the above knowledge, the {\em voltage drop} across the desired system {$(R)$ in Figure} \ref{fig:sysIDgeneric} can be computed.
 This method allows to develop an observation model that is based on the current $i(k)$ and $v(k)$, potentially resulting in high SNR of the observation model. 
 However, this approach depends on the accuracy of the model assumed in \eqref{eq:combined} which may be susceptible to changes due to aging, temperature and use patterns.  
\end{itemize}

\begin{figure}
\begin{center}
\subfloat[][Measurement setup]
{\includegraphics[width=.35\columnwidth]{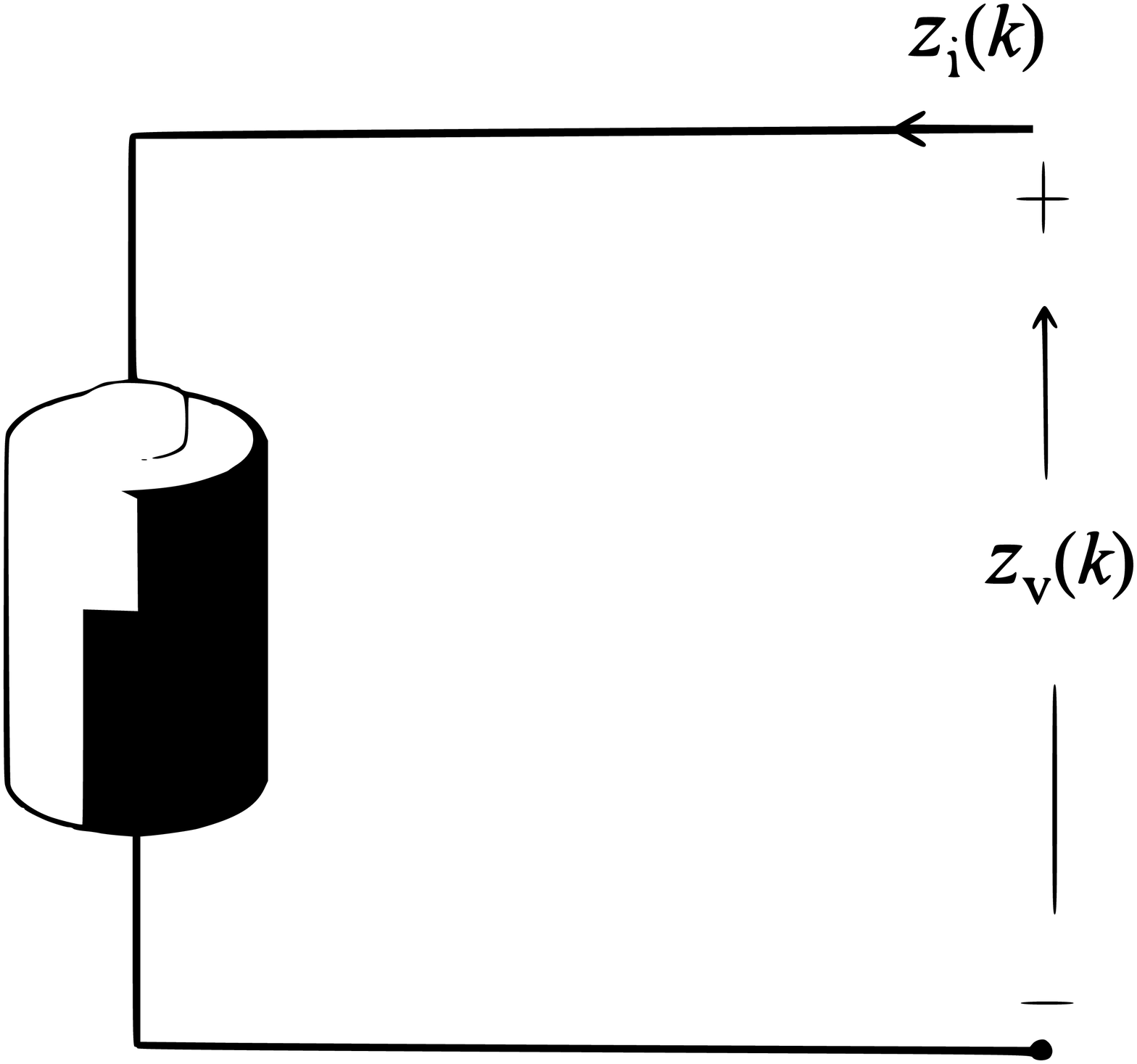}}
\subfloat[][Generic equivalent circuit model]
{\includegraphics[width=0.7\columnwidth]{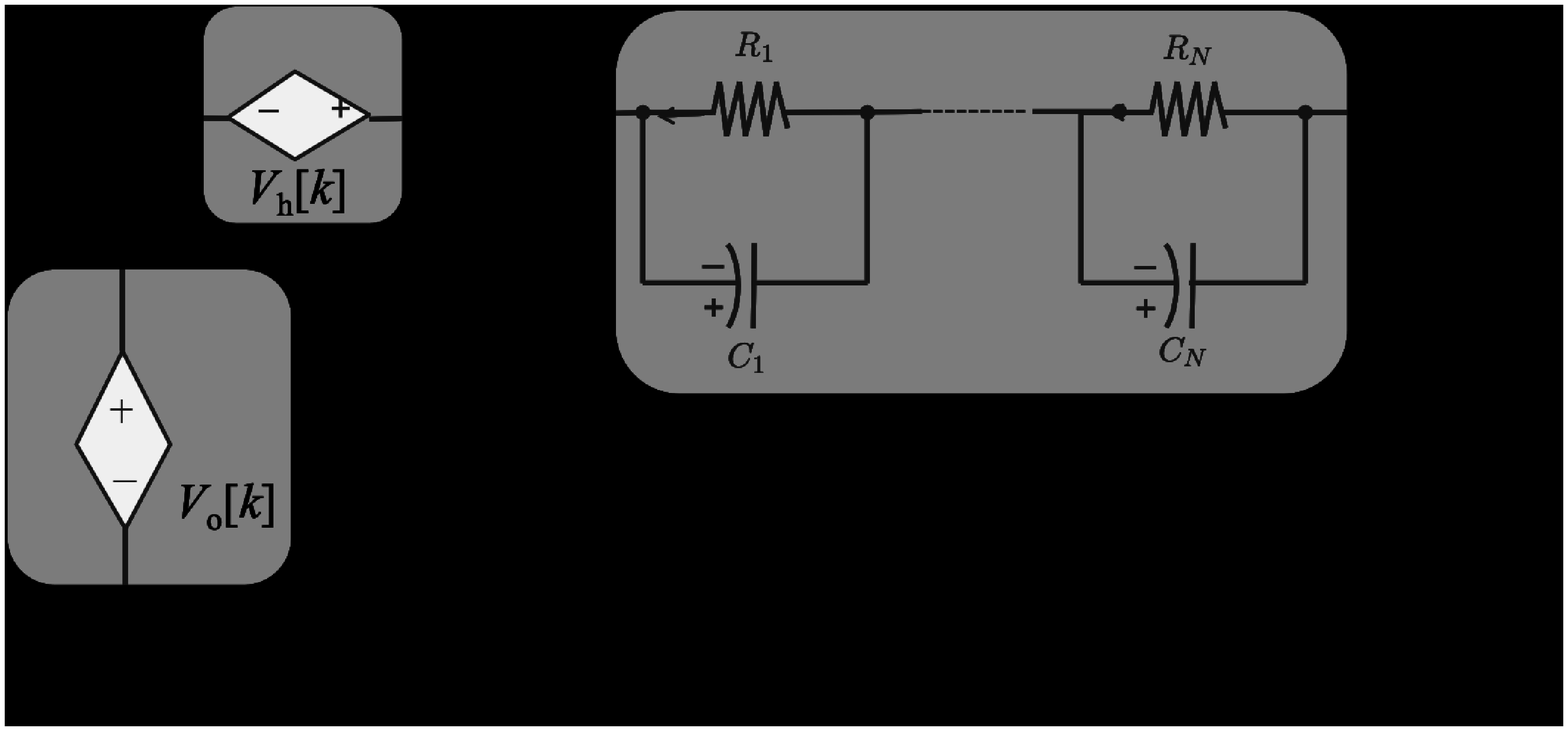}\label{genericmodel}}
\end{center}
\caption{
{\bf Experiment using realistic battery data.}
The subfigure (a) shows the measurement and subfigure (b) shows a generic equivalent circuit model of a battery. 
}
\label{fig:modelN}
\end{figure}

The battery ECM model shown in Figure \ref{fig:sysIDgeneric}(b) is a simplistic one.
{However, this model can accurately represent the battery behavior when there is constant charge/discharge current going through it.
A good example of this scenario is when a battery is being charged using CC-CV or when an electric vehicle is moving with constant velocity on a highway.}
A more detailed version of a typical battery ECM is shown in Figure \ref{fig:modelN}\subref{genericmodel}, which includes the hysteresis and relaxation mechanisms of a battery in addition to the series internal resistance -- indeed, similar model identification is common in many other applications \cite{ferrigno2008measurement,guha2018online,couraud2017real}. 
{A more detailed ECM can result in higher estimation accuracy but they require higher computational complexity, in \cite{zou2018review} a review of different models that can be used is provided and \cite{tang2019load} proposes a way to switch between different models based on the current scenario.}
Similar to before, the ECM identification can be approached in two different premises:
(a) By considering the unknown system as a blackbox and 
(b) By assuming the knowledge of the unknown system (estimating the model parameters, if needed). 
Numerous approaches have been presented in the past for ECM identification of Li-ion batteries including those presented in \cite{BFGpart12014} \& \cite{ArifEPEC2019}. 
However, the discussions presented in this paper warrants a second look at the existing algorithms for ECM parameter identification in rechargeable batteries. 
{For future work the proposed approach will be derived and applied to higher order models to see the effects of taking the voltage and current noise in consideration on the parameter estimation and state of charge estimation accuracy.} 

\section{Conclusions} 
\label{sec:conclusions}

In this paper, we considered the problem of electrical equivalent circuit model (ECM) parameter estimation based on noisy voltage and current measurements. 
First, we demonstrated the challenges faced by traditional least squares (LS) estimation approach in estimating the resistance based on measured voltage and measured current --- it was shown that the LS estimates are biased when the model uncertainty, i.e., the noise in current, is significant.
To this problem, we demonstrated a solution based on the total least squares (TLS) approach and demonstrated that the TLS estimates are indeed nearly unbiased and efficient under practical assumptions of varying SNR. 
Further, we discussed the need to implement ECM parameter identification in recursive manner for computational efficiency and to accommodate the possibility of time-varying ECM parameters. 
Later, a recursive implementation of the TLS approach is presented based on a forgetting factor; it was then shown that the convergence of the recursive TLS is sensitive to the forgetting factor in terms of the variance of the recursive TLS estimates and time to converge.  
In order to retain the benefits of the TLS estimator, a Kalman filter is proposed. 
The proposed filter, named total Kalman filter (TKF), is used to smooth the variance found in the TLS estimator;
the proposed TKF estimator is shown to be both unbiased and efficient in practical operational ranges.

\section*{Acknowledgements} 
B. Balasingam acknowledges the support of the Natural Sciences and Engineering Research Council of Canada (NSERC) for
financial support under the Discovery Grants (DG) program [funding reference number RGPIN-2018-04557].
B. Balasingam acknowledges the help of Mostafa Ahmed and Arif Raihan for their help searching through relevant literature for Section I of this manuscript. 
Research of K. Pattipati was supported in part by the U.S. Office of Naval Research and US Naval Research Laboratory under Grants
\#N00014-18-1-1238, \#N00173-16-1-G905, \#HPCM034125HQU and by a Space Technology Research Institutes grant (\#80NSSC19K1076) from NASA's Space Technology Research Grants Program.

\end{document}